\documentclass[11pt]{article}
\pdfoutput=1
\usepackage{jcapmod}
\usepackage{booktabs}
\usepackage[english]{babel}
\usepackage{amsmath,amssymb,amsbsy,amstext, amsthm, simplewick}
\usepackage{wrapfig}
\usepackage{hyperref}
\usepackage{graphicx}
\usepackage{amsfonts}
\usepackage{amssymb}
\usepackage{subfig}
\usepackage{upgreek}
 \usepackage{exscale,relsize}
 \usepackage[makeroom]{cancel}
\usepackage{soul}
\usepackage{bbold}
\usepackage{lipsum}
\usepackage{mdframed}
\usepackage{mathtools}
\usepackage{tikz-cd}
\usepackage[export]{adjustbox}
\usepackage[outdir=./]{epstopdf}

\usetikzlibrary{shapes,arrows}

\RequirePackage{color}

\usepackage{colortbl}
\definecolor{rp}{cmyk}{0.2, 1, 0.6, 0}
\definecolor{green2}{cmyk}{0, 1, 0.5, 0}
\definecolor{lightgreen}{cmyk}{0.2, 0, 0.2, 0.2}
\definecolor{lightgray}{cmyk}{0.1,0.2,0,0.1}
\definecolor{lightgray2}{cmyk}{0.4,0.4,0,0.8}
\definecolor{black}{cmyk}{1.0,1.0,1.0,1.0}

\allowdisplaybreaks[1]

\usepackage{colortbl}
\definecolor{lightgreen}{cmyk}{0.2, 0, 0.2, 0.2}
\definecolor{lightgray}{cmyk}{0.1,0.2,0,0.1}
\definecolor{lightgray2}{cmyk}{0.1,0.1,0,0.1}

\makeatletter
\newlength{\apb@width}
\newcommand{\autoparbox}[2][c]{\settowidth{\apb@width}{#2}\parbox[#1]{\apb@width}{#2}}

\newcommand{\Cen}[2]{%
  \ifmeasuring@
    #2%
  \else
    \makebox[\ifcase\expandafter #1\maxcolumn@widths\fi]{$\displaystyle#2$}%
  \fi
}
\makeatother

\numberwithin{equation}{section}

\def\Tr{{\rm Tr}}

\def\beq{\begin{equation}}
\def\eeq{\end{equation}}
\def\bea{\begin{eqnarray}}
\def\eea{\end{eqnarray}}

\def\Beq{\begin{equation}\begin{aligned}}
\def\Eeq{\end{aligned}\end{equation}}

\def\bk{\vec{k}}
\def\bx{\vec{x}}

\def\blambda{{\boldsymbol{\lambda}}}
\def\bomega{{\boldsymbol{\omega}}}
\def\Ns{{N_{\rm s}}}
\def\Nf{N_{\rm f}}

\def\U{{\sf U}}

\def\G{{\sf G}}
\def\F{{\sf F}}
\def\Q{{\sf Q}}

\def\hvp{\delta\varphi}
\def\M{{\sf M}}
\def\R{{\sf R}}

\def\f{{\sf f}}
\def\g{{\sf g}}
\def\n{{\sf n}}

\def\p{{\sf p}}

\def\u{{\sf u}}
\def\v{{\sf v}}

\def\f{{\sf f}}
\def\tf{{\tilde{\sf f}}}
\def\L{{\sf \Lambda}}
\def\m{{\sf m}}
\def\Nf{{N_{\rm{f}}}}

\def\trasp{\mathsf{T}}
\def\I{\mathbb{1}}
\def\hth{\delta\theta}

\def\hph{\delta\phi}
\def\hvp{\delta\varphi}
\def\Fo{\mathcal{F}}
\newcommand{\bsf}{\bf\sf}
\DeclareRobustCommand{\SkipTocEntry}[4]{}

\begin{document}

\hypersetup{pageanchor=false}

\begin{titlepage}

\setcounter{page}{1} \baselineskip=15.5pt \thispagestyle{empty}

\bigskip\

\vspace{1cm}
\begin{center}

{\fontsize{20.74}{24}\selectfont  \sffamily \bfseries  Multifield Stochastic Particle Production: Beyond a Maximum Entropy Ansatz}

\end{center}

\vspace{0.2cm}

\begin{center}
{\fontsize{12}{30}\selectfont  Mustafa A.~Amin$^{\clubsuit}$\footnote{mustafa.a.amin@gmail.com}, Marcos A.~G.~Garcia$^{\clubsuit}$\footnote{marcos.garcia@rice.edu}, Hong-Yi~Xie$^{\blacklozenge,\clubsuit}$ and Osmond Wen$^{\clubsuit}$}
\end{center}

\begin{center}

\vskip 7pt

\textsl{$^\clubsuit$ Department of Physics \& Astronomy, Rice University, Houston, Texas 77005-1827, U.S.A.}
\vskip 7pt

\textsl{$^ \blacklozenge$ Department of Physics, University of Wisconsin-Madison, Madison, Wisconsin 53706, U.S.A.}
\vskip 7pt



\end{center}

\vspace{1.2cm}
\hrule \vspace{0.3cm}
\noindent {\sffamily \bfseries Abstract} \\[0.1cm]
We explore non-adiabatic particle production for $\Nf$ coupled scalar fields in a time-dependent background with stochastically varying effective masses, cross-couplings and intervals between interactions. Under the assumption of weak scattering per interaction, we provide a framework for calculating the typical particle production rates after a large number of interactions. 
After setting up the framework, for analytic tractability, we consider interactions (effective masses and cross couplings) characterized by series of Dirac-delta functions in time with amplitudes and locations drawn from different distributions. Without assuming that the fields are statistically equivalent, we present closed form results (up to quadratures) for the asymptotic particle production rates for the $\Nf=1$ and $\Nf=2$ cases. We also present results for the general $\Nf >2$ case, but with more restrictive assumptions. We find agreement between our analytic results and direct numerical calculations of the total occupation number of the produced particles, with departures that can be explained in terms of violation of our assumptions. 

We elucidate the precise connection between the maximum entropy ansatz (MEA) used in Amin \& Baumann (2015) and the underlying statistical distribution of the self and cross couplings.  We provide and justify a simple to use (MEA-inspired) expression for the particle production rate, which agrees with our more detailed treatment when the parameters characterizing the effective mass and cross-couplings between fields are all comparable to each other. However, deviations are seen when some parameters differ significantly from others. We show that such deviations become negligible for a broad range of parameters when $\Nf\gg 1$.  

\vskip 10pt
\hrule
\vskip 10pt

\vspace{0.6cm}
 \end{titlepage}

\hypersetup{pageanchor=true}

\tableofcontents

\newpage

\section{Introduction}
Repeated bursts of particle production are possible during inflation and reheating after inflation. Such events typically result from a non-adiabatic change in the effective mass or the couplings between fields as the background evolution of the fields passes through special points in field space. For example, sharp turns in field trajectories or certain fields becoming effectively massless can result in significant particle production. The produced particles can impact curvature fluctuations and/or cause a change in the way energy is transferred between the effective inflaton field and the spectator/daughter fields (see for example \cite{Kofman:1994rk,Shtanov:1994ce,Kofman:1997yn,Tye:2008ef,Ashoorioon:2008qr,Green:2009ds,Tye:2009ff,Braden:2010wd, Achucarro:2010jv, Achucarro:2010da, Chen:2011zf, Battefeld:2011yj, Dias:2012nf, McAllister:2012am, Battefeld:2012wa, Greenwood:2012aj, Marsh:2013qca,Easther:2013rva,Hertzberg:2014iza,Watanabe:2015eia,Dias:2015rca,Figueroa:2015rqa,Jain:2015mma,DeCross:2015uza,Chluba:2015bqa,Amin:2014eta,Barnaby:2012xt,Lozanov:2016hid,Pearce:2016qtn,Flauger:2016idt,Pearce:2017bdc}).

For a fully specified model which includes all the interaction terms between fields, one can always numerically compute the background evolution of the fields. Subsequently, one can calculate particle production in the fields for any finite wavenumber by solving the mode equation in this background. Such fully specified models are rarely available. Even for a fully specified model with many fields there might be many background solutions. Since we might not know a-priori which trajectory is taken, it is not always the most efficient way to go carry our particle production calculations (however, see for example, \cite{Price:2014xpa,Dias:2016rjq} for some recent numerical approaches in calculating multifield inflationary perturbations).

An alternate approach is to think about particle production in a statistical ensemble of models or trajectories within a model, and focus on calculating the typical particle production rate. This is the approach we take in this paper. The work here is heavy up front, in terms of setting up the framework, but once set up it can be applied to a wide range of models where only statistical information about the parameters is available, and only coarse-grained information about the final observables is needed. Given the complexity of ultraviolet completion of the models of inflation, and the simplicity of the data from the early universe we believe that this statistical approach is a reasonable one to pursue. 

An earlier paper \cite{Amin:2015ftc} by one of us took advantage of a mathematical mapping between current conduction in wires with impurities and particle production in a time dependent background. In that work, a significant simplification in the results is seen due to a {\it{maximum entropy ansatz}} (MEA), which heuristically assumes statistically equivalent fields.  In this paper, we take a more detailed approach which allows us to go beyond statistically equivalent fields. However, the cost is an increase in complexity of our derivations. Nevertheless, once obtained, our results are relatively simple to understand and use, and do reduce to the results of \cite{Amin:2015ftc} in their overlapping domain of validity (which we delineate). 

In this paper, we first provide a more explicit connection between a microscopic (but still statistical) description of the scatterers and the MEA. Without relying on this ansatz, we derive a more general Fokker-Planck equation for the time dependent probability distribution of the occupation number of fields. The equation is valid under the assumption that the particle production for each interaction is small. Using this equation, we calculate the typical total occupation number of the fields. For the sake of analytic tractability, each interaction is modeled by a Dirac-delta function in time whose amplitude and location is drawn from different distributions. We are able to derive closed form (up to quadratures) results for the typical total occupation number after a large number of scatterings $\Ns\gg 1$, when the number of fields $\Nf=1$ and $\Nf=2$. For $\Nf>2$, the derivation becomes increasingly complex. By making additional technical assumptions regarding the relative strengths of the effective masses and cross couplings of the fields, we are able to calculate the total occupation number for an arbitrary number of fields. As might be expected, we see a significant simplification in our results for $\Nf\gg 1$. In general, for simple results, it is essential to have $\Ns\gg 1$ as well as $\Nf=1,2$ or $\Nf\gg 1$.


We briefly discuss some important assumptions made in this paper, which we will also reiterate in the conclusions. First, we consider particle production in Minkowski space rather than an expanding universe which would be more appropriate for inflation and reheating applications. Part of the effect of expansion is accounted for by randomizing the location of the events -- this mimics the phase scrambling that takes place in an expanding universe \cite{Kofman:1997yn}, but a more careful treatment of the effects due to expansion is left for future work. Second, we treat each Fourier mode independently; it is a linear treatment about a spatially homogeneous background. As significant particle production takes place, this assumption might be broken. Thus, our results are only valid as long as the linearized equations hold. Finally, we assume that the particle production per event is weak, and the number of particle production events is large for the occupation number to build up. More technical assumptions are discussed in the following sections once the appropriate terminology has been introduced.

Without any attempts at being exhaustive, we would like to highlight a few papers that provide some background and context for our work. Non-perturbative particle production in the context of reheating has a long history \cite{Kofman:1994rk,Shtanov:1994ce} (see \cite{Amin:2014eta} for a review). Similarly, in the context of inflation, see for example \cite{Green:2009ds, Barnaby:2012xt, Flauger:2016idt,Pearce:2017bdc} (also see the review \cite{Chluba:2015bqa}). In \cite{Bassett:1997gb, Zanchin:1997gf,Zanchin:1998fj}, the authors took advantage of the connection between stochastic particle production and random Schr\"odinger operators in the context of reheating. The implications of weak disorder in the early universe and inflationary perturbations was discussed in \cite{Green:2014xqa}. In \cite{Amin:2015ftc}, a statistical framework for calculating non-adiabatic particle production rates in stochastically coupled (but statistically equivalent) multi-field scenario was presented, which in turn took advantage of related existing techniques used to understand current conduction in disordered wires \cite{anderson1958absence,muller2010disorder,Beenakker:1997zz,mello2004quantum}.

The rest of the paper is organized as follows. In section \ref{sec:TheModel}, we provide a Lagrangian for the perturbations in the fields undergoing non-adiabatic particle production. In section \ref{sec:TMA}, we discuss the {\it Transfer Matrix Approach} where we provide the formal expressions for the occupation number of the fields in terms of a product of transfer matrices evaluated at each non-adiabatic event. Under the assumption of weak particle production per event, we derive a general Fokker-Planck equation describing the evolution of the probability density of the parameters characterizing the particle production events. In section \ref{sect:MEA}, we connect the MEA to the statistical description of the underlying model. In section \ref{sect:onefield}, we consider the single field ($\Nf=1$) case as a warm-up example. Moving beyond MEA, in section \ref{sect:twofield}, we consider the $\Nf=2$ case in detail. We first calculate the coefficients of the Fokker-Planck equation by taking averages over the properties of the underlying non-adiabatic events. We then calculate the typical occupation number of the field using the Fokker-Planck equation. In section \ref{sect:manyfields}, we generalize to the $\Nf>2$ case, and again compare with the corresponding results in \cite{Amin:2015ftc}. In section \ref{sect:summary}, we discuss our results, reiterate the assumptions and caveats relevant for the results, and outline our plans for future work. A number of technical details are relegated to the appendices. In Appendix \ref{ap:FPd}, we give details of the derivation of the multi-field Fokker-Planck equation. In Appendix \ref{ap:nf2}, we provide some of the explicit calculations for the coefficients in the $\Nf=2$ case. In Appendix \ref{ap:nm}, we show the explicit numerical algorithm used for calculating the particle production rate. Finally, in Appendix \ref{ap:Fav} we discuss technical details of the Haar measure for the transfer matrices.


\subsubsection*{Notation and Conventions}
Throughout, we will use 
natural units, $c=\hbar \equiv 1$.
The time variable will be $\tau$, and overdots will denote derivatives with respect to $\tau$.
We will use a bold {\sf sans-serif font} for matrices, e.g.~$\M, \n, \blambda$. The indices $a,b,\ldots$ will be used for field indices. 
We use the $+---$ metric convention.
\section{The Model}
\label{sec:TheModel}
Consider $\Nf$ coupled scalar fields, $\chi^a(\tau,{\bx})$, with masses $m_a$ and time-dependent stochastic masses and couplings $m^{\rm s}_{ab}(\tau)$. Here $a,b=1,2,\hdots, \Nf$. The quadratic action for these fields is taken to be
\Beq
S^{(2)}=\int d^4 x&\,\mathcal{L}=\int d^4 x\sum_{a,b=1}^{\Nf}\left(\frac{1}{2}\delta_{ab}\partial_\mu\chi^a\partial\chi^b-\frac{1}{2}\mathcal{M}_{ab}(\tau)\chi^a\chi^b\right)\,,\\
&\mathcal{M}_{ab}(\tau)=m_a^2\delta_{ab}+m^{\rm s}_{ab}(\tau)\,.
\Eeq
The above action is to be interpreted as the action for perturbations around a complicated time dependent background due to the presence of many fields.  The equations of motion for the fields $\chi^a$ are
\Beq
\left(\frac{d^2}{d \tau^2}-\nabla^2+m_a^2\right)\chi^a(\tau,{\bx})+\sum_{b=1}^{\Nf}m^{\rm s}_{ab}(\tau)\chi^b(\tau,{\bx})=0\,.
\Eeq
The equations of motion of the Fourier modes of the field $\chi^a(\tau,\bk)$ are
 \Beq
&\left(\frac{d^2}{d \tau^2}  + \omega_a^2\right)\chi^a(\tau,\bk)+\sum_{b=1}^{\Nf}m^{\rm s}_{ab}(\tau)\chi^b(\tau,\bk)=0\,,\\
&\omega_a^2(k)=k^2+m_a^2\,,
\Eeq
where we have used the symbol $\chi^a$ for the Fourier transform of the fields as well as the fields themselves. We will be explicitly dealing with Fourier space quantities from now on. To reduce clutter, the dependence of the fields and $\omega_a$ on the wavenumber $k$ will be suppressed.

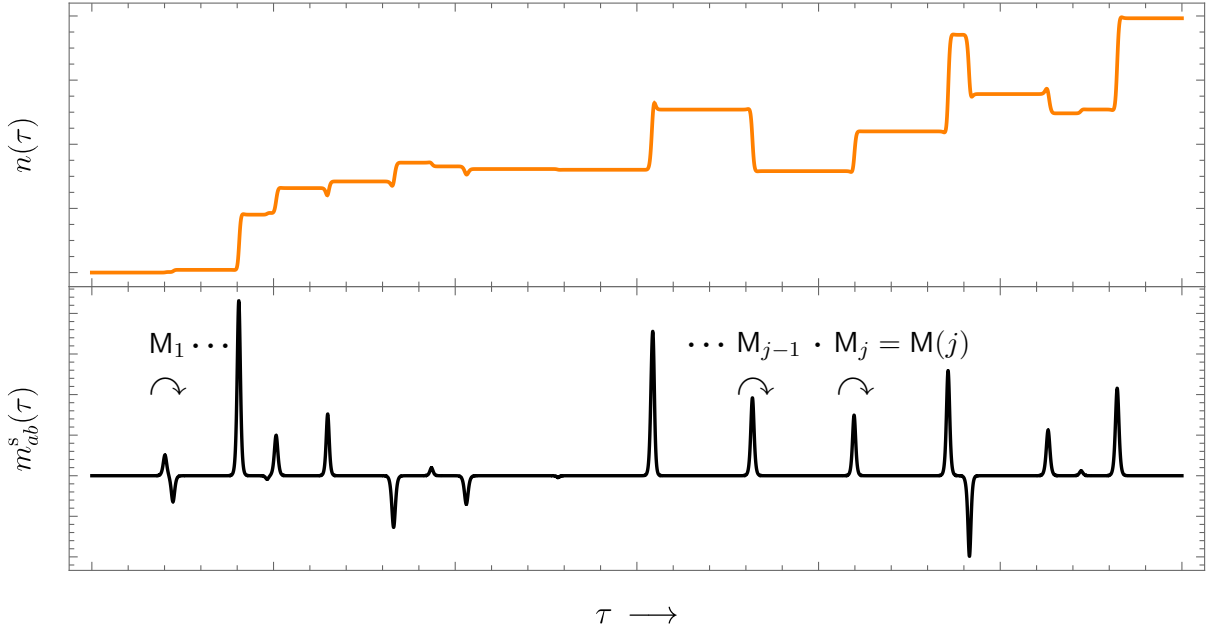
\begin{figure}[h!] 
   \centering
   \centering
   \begin{tikzpicture}
    \node[anchor=south west,inner sep=0] (image) at (0,0) {\includegraphics[width=\textwidth]{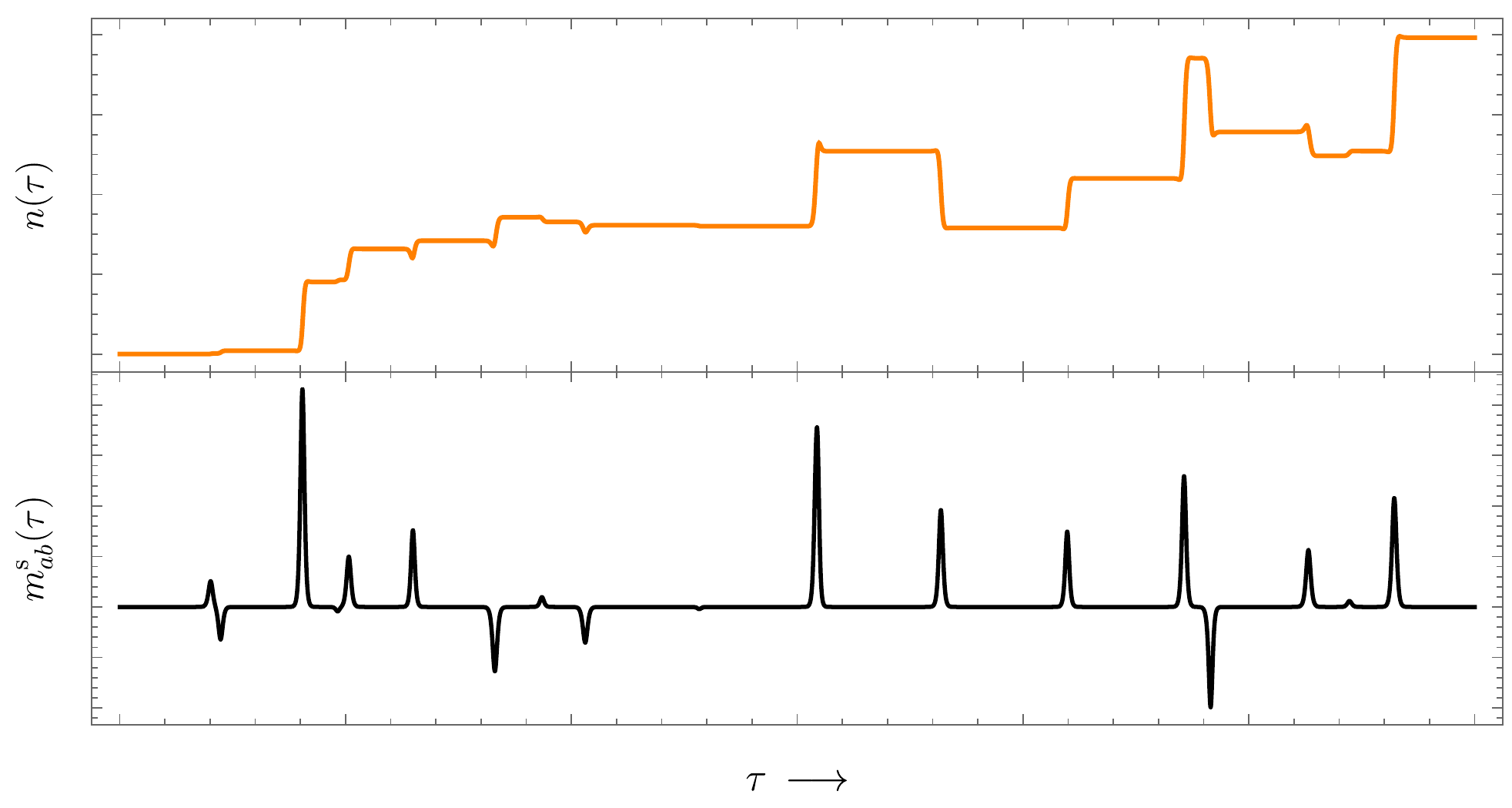}};
    \begin{scope}[x={(image.south east)},y={(image.north west)}]
    \node at (0.685,0.45) {$\boldsymbol{\cdots}\,\M_{j-1}\,\boldsymbol{\cdot}\,\M_j = \M(j)$};
    \node at (0.162,0.45) {$\M_1\,\boldsymbol{\cdots}$};
    \node at (0.625,0.385) {\Large $\curvearrowright$};
    \node at (0.707,0.385) {\Large $\curvearrowright$};
    \node at (0.142,0.385) {\Large $\curvearrowright$};
   \end{scope}%
   \end{tikzpicture}%
   \caption{Lower panel: The stochastic masses and cross couplings $m_{ab}^s(\tau)$. Upper panel: The occupation number for a field. The non-adiabatic interactions result in a random walk-like behavior (with drift) for the occupation number $n$. Note that the change in occupation number will depend on the wavenumber $k$.}
   \label{fig:HillsValleys}
\end{figure}

The time dependence of $m^{\rm s}_{ab}(\tau)$ can be quite complex. For example, $m^{\rm s}_{ab}(\tau)$ can consist of a series of well separated ``hills" and ``valleys", uniformly distributed in time. Between these hills and valleys, the fields are assumed to be free and uncoupled (see fig.~\ref{fig:HillsValleys}).

\section{The Transfer Matrix Approach}\label{sec:TMA}
The Fourier mode of the field $\chi^a$ after the $j$-th non-adiabatic event is
\beq
\chi^a_{j}(\tau)\equiv\frac{1}{\sqrt{2\omega_a}}\left[\beta^a_{j} e^{i\omega_a\tau}+\alpha^a_{j} e^{-i\omega_a\tau}\right] , \label{equ:chi}
\eeq
where the overall normalization is chosen for future convenience.  The Wronskian of the solutions $\chi^a_j$ and ${\chi^a_j}^*$ is a constant, $W[\chi^a_j, {\chi^a_j}^*] \equiv 1$, which in terms of the coefficient in eq.~\eqref{equ:chi} implies $|\alpha^a_j|^2-|\beta^a_j|^2=1$. 

The coefficients $\vec{\beta}_j=(\beta^1_j,\beta^2_j\hdots\beta^\Nf_j)^T$ and $\vec{\alpha}_j=(\alpha^1_j,\alpha^2_j\hdots\alpha^\Nf_j)^T$ before and after the $j$-th scattering, are related by the transfer matrix $\M_j$.
\begin{equation}
\begin{pmatrix} \vec{\beta}_{j} \\ \vec{\alpha}_{j} \end{pmatrix}= 
 {\M}_j
  \begin{pmatrix} \vec{\beta}_{j-1} \\ \vec{\alpha}_{j-1} \end{pmatrix}\, . \label{equ:transfer2}
\end{equation}
For a single scattering we can always write $\M_j=\I+\m_j$ where $\m_j$ contains the necessary information about the scattering (for weak scattering $|\m_j|\ll 1$). We can connect the coefficients before and after $j$ scatterings by simply chaining together the transfer matrices
\begin{equation}\label{eq:complaw}
\begin{pmatrix} \vec{\beta}_{j} \\ \vec{\alpha}_{j} \end{pmatrix}= 
 {\M}(j)
  \begin{pmatrix} \vec{\beta}_{0} \\ \vec{\alpha}_{0} \end{pmatrix}\,,\qquad \textrm{where}\qquad\M(j)=\M_{j}\M_{j-1}\hdots\M_1.
\end{equation}
Note that $\M(j)$ denotes the transfer matrix of $j$ scatterings, whereas $\M_j$ is the transfer matrix for the $j$-th scattering.
\subsection{The Occupation Number}
We define the {\it occupation number} of each field after $j$ scatterings by\footnote{We note that this definition agrees with the usual one when thinking about $\chi^a(\tau,\vec{k})$ as mode functions for a single field undergoing adiabatic evolution. In more general time dependent backgrounds, and with coupled fields, the connection is not always so simple. See for example \cite{Mukhanov:2007zz, Amin:2014eta,DeCross:2015uza}.}
\Beq
&n_a(j)=\frac{1}{2\omega_a}\left(|\dot{\chi}^a_j|^2+\omega_a^2|{\chi}^a_j|^2\right)-\frac{1}{2}=|\beta^a_j|^2\,,
\Eeq
and the sum of the occupation numbers of all the fields is defined as
\Beq
n(j)\equiv \sum_{a=1}^{\Nf}n_a(j)\,.
\Eeq
For technical reasons which will become obvious towards the end of the calculation, the total occupation number is the quantity we will focus on. A useful identity relating the total occupation number defined above and the transfer matrix is given by
\Beq
n(j)=\frac{1}{4}\Tr\left[\M(j)\M^\dagger(j)-\I\right]\,.
\Eeq
The above expression for the total occupation number motivates the following definition:
\Beq
\R_j\equiv  \M_j\M_j^\dagger\,,\qquad\textrm{and correspondingly}\qquad \R(j)\equiv\M(j)\M^\dagger(j)\,.
\Eeq
For the total occupation number, we only care about $\R$ and not $\M$. This simplifies matters considerably. Note that $\R$ is Hermitian. The total occupation number is related to the sum of the eigenvalues of $\R$. 

\subsection{Parametrizing \texorpdfstring{$\R$}{R}}

How many parameters $\{\lambda_a\}$ are needed to describe $\R$? A general parametrization of the $\R$ matrix is
\Beq
\R = \underbrace{\begin{pmatrix} \u & 0 \\ 0 & \u^\ast \end{pmatrix}}_{\U}
               \underbrace{\begin{pmatrix} \f &  \tf \\ \tf & \f \end{pmatrix}}_{\F}
               \underbrace{\begin{pmatrix} \u^\dagger & 0 \\ 0 & \u^\trasp \end{pmatrix}}_{\U^\dagger}\,=\U \F \U^\dagger,
\Eeq 
where $\tf\equiv \sqrt{\f^2-1}$. In the above expression $\f$ is a diagonal matrix consisting of $\Nf$ eigenvalues and $\u$ is a $\Nf\times\Nf$ unitary matrix parametrized by $\Nf^2$ angular variables. Note that $\U$ and $\F$ are $2\Nf\times2\Nf$ matrices with $\Nf\times \Nf$ blocks constructed out of $\f$ and $\u$.  Thus, $\R$ is parametrized by $\Nf^2+\Nf$ total variables. This should be contrasted with $\M$, for which a general parametrization has the form~\cite{mello1988macroscopic}
\beq\label{eq:Mparg}
\M =\begin{pmatrix} \u & 0 \\ 0 & \u^\ast \end{pmatrix}
               \begin{pmatrix} \sqrt{1+\n} &  \sqrt{ \n} \\ \sqrt{ \n} & \sqrt{1+\n} \end{pmatrix}
               \begin{pmatrix} \v & 0 \\ 0 & \v^\ast \end{pmatrix}\,,
\eeq
where $ \n\equiv (\f-\I)/2$, and where $\v$ is a $\Nf\times\Nf$ unitary matrix parametrized by $\Nf^2$ angular variables, for a total of $2\Nf^2+\Nf$ variables characterizing $\M$.  The parameters of $\R$ will be denoted by $\{\lambda_a\}=\{f_1,\hdots,f_{\Nf},\theta_1,\hdots,\theta_{\Nf^2}\}$ where $f_a$ are the diagonal entries of $\f$ and $\theta_a$ are the angular variables for the unitary matrix $\u$. In this parametrization, note that
\Beq\label{eq:njfa}
n(j)=\frac{1}{4}\Tr[\R-\I]=\frac{1}{2}\Tr[\f(j)-\I]=\frac{1}{2}\sum_{a=1}^\Nf (f_a-1)\,.
\Eeq

\newpage
\subsection{The Fokker Planck Equation}
Our next goal is to figure out how $\R$ evolves as the fields experience scattering events. Assuming that the parameters describing the scatterers are drawn from some distribution, we wish to derive an equation for the probability density $P_{\tau}(\R)$. This can then be used to get the expectation value of the occupation number (or any other function depending on variables in $\R$). 

We start by adding a small time interval $\delta \tau$ with a single scatterer to an existing interval $\tau$ with $j$ scattering events. We will consider the case that the scatterer is weak. That is, the change in $\M$ is small due to the additional interval. The transfer matrix $\M_{\tau+\delta\tau}$ for the elongated interval can be written in terms of $\M_1\equiv \M_{\tau}$ and $\M_2\equiv \M_{\delta\tau}$ as $\M_{\tau+\delta\tau}=\M_{\delta\tau}\M_{\tau}$ due to the composition law in eq.~\eqref{eq:complaw}. As the underlying processes occurring in the $\tau$ and $\delta\tau$ strips are assumed to be uncorrelated, the probability densities $P_{\tau}(\M_1)$ and $P_{\delta\tau}(\M_2)$ are statistically independent. One can then show that the density $P_{\tau+\delta\tau}(\M)$ can be obtained as the convolution~\cite{mello2004quantum}
\beq\label{eq:smoluchowsky}
P_{\tau+\delta\tau}(\M)=\int d\mu(\M_2)\,P_{\tau}(\M_2^{-1}\M)P_{\delta\tau}(\M_2)\,,
\eeq
where $d\mu(\M)$ denotes the invariant (Haar) measure with respect to the transfer matrix group, with the property that $d\mu(\M)=d\mu(\M_0\M)$ with $\M_0$ being any other (fixed) transfer matrix.

Equation~(\ref{eq:smoluchowsky}) corresponds to the Smoluchowsky equation for the present (Markovian) process. As it is well known from the theory of Brownian motion, the Smoluchowsky equation implies that the probability distribution of the transfer matrix parameters, and consequently that for the $\R$-matrix parameters $P_{\tau}(\{\lambda_a\})$, evolves with time according to the (forward) Kramers-Moyal expansion
\Beq\label{eq:FPlambda}
\partial_\tau P&=-\sum_{b=1}^{\Nf^2+\Nf} \partial_{\lambda_b}\left[\frac{{\langle \delta\lambda_b\rangle}_{\delta\tau}}{\delta\tau} P\right]+\frac{1}{2!}\sum_{b,c=1}^{\Nf^2+\Nf}\partial_{\lambda_b}\partial_{\lambda_c}\left[\frac{{\langle \delta\lambda_b\delta\lambda_c\rangle}_{\delta\tau}}{\delta\tau} P\right]+\hdots
\Eeq
The $\delta\lambda_a$ in the above expression should be thought of as the ``small" increment of the parameters $\lambda_a$ (relative to their values {\em before} the addition of the interval $\delta\tau$) due to a single additional scatterer in the time interval $\delta\tau$ (weak scattering). The expectation value is over the probability distribution describing the properties of the scatterer in the interval $\delta\tau$. For a proof of the expansion in eq.~\eqref{eq:FPlambda} we refer the reader to Appendix~\ref{ap:FPd}. In the weak scattering limit, the Kramers-Moyal expansion can be curtailed at second order, and eq.~\eqref{eq:FPlambda} may be referred to as the Fokker-Planck (FP) equation for $P_{\tau}(\{\lambda_a\})$.\footnote{In the single field limit it can be explicitly shown that $\langle \delta n\rangle = (\mu\,\delta\tau)(1+2n)$, $\langle(\delta n)^2\rangle = (\mu\,\delta\tau)2n(1+n)+\mathcal{O}[(\mu\,\delta\tau)^2]$, $\langle(\delta n)^{2k}\rangle = \mathcal{O}[(\mu\,\delta\tau)^k]$ and $\langle(\delta n)^{2k+1}\rangle = \mathcal{O}[(\mu\,\delta\tau)^{k+1}]$ for $k>1$, where $\mu$ denotes the local mean particle production rate~\cite{Amin:2015ftc}. }

It is often not necessary to get an explicit solution for $P$. Instead we can integrate eq.~\eqref{eq:FPlambda} to obtain equations for the expectation values of quantities of interest. Let us consider a function $F(n)$ where $n$ is the total occupation number of the fields. Note that $n=(1/2)\sum_{a=1}^{\Nf}(f_a-1)$ only depends on the $\Nf$ eigenvalues $\{f_a\}$ and not on the angular variables. Multiplying eq.~\eqref{eq:FPlambda} with $F(n)$ and integrating over all $\Nf^2+\Nf$ variables we arrive at
\Beq\label{eq:dtfn}
\partial_\tau \langle F(n)\rangle=\sum_{a=1}^\Nf\left \langle\frac{F'(n)}{2}\frac{\langle{\delta f_a}\rangle_{\delta\tau}}{\delta\tau}+\sum_{b=1}^{\Nf}\frac{F''(n)}{8}\frac{\langle{\delta f_a\delta f_b}\rangle_{\delta\tau}}{\delta\tau}\right\rangle\,,
\Eeq
where $\langle \hdots \rangle=\int \prod_{a=1}^{\Nf^2+\Nf} d\lambda_a P_{\tau}(\{\lambda_a\})(\hdots) $. In deriving the above expression we have repeatedly used integration by parts, dropped the boundary terms, taken advantage of the fact that the $\Nf^2$ variables of $\u(\{\theta_\alpha\})$ form a compact manifold, and that $F(n)$ is a function of the $\Nf$ eigenvalues $\{f_a\}$ only. We have also used the expression for $n$ in terms of the eigenvalues $\{f_a\}$. Note in particular that the sums are over $\Nf$ variables only, however the integration is over all $\Nf^2+\Nf$ variables. Moreover the coefficients $\langle\hdots\rangle_{\delta\tau}$ as well as $P$ might be non-factorizable functions of all $\Nf^2+\Nf$ variables. 

Our interest will be focused on the so-called typical occupation number, defined as
\beq\label{eq:ntypd}
n_{\rm typ} \equiv \exp[\langle\ln(1+n)\rangle] - 1\,.
\eeq
The quantity $n_{\rm typ}$ is a good measure of the typical number of particles produced, as its variance grows slower than the square of the mean. For the mean occupation number $\langle n\rangle$ the ratio $(\langle n^2\rangle -\langle n\rangle^2)/\langle n\rangle^2$ grows with time (see the discussion in  \cite{Amin:2015ftc}).

With $n_{\rm typ}$ in mind, consider the function $F(n)=\ln(1+n)$. Due to the relation (\ref{eq:njfa}), eq.~(\ref{eq:dtfn}) can be then be written as
\Beq\label{eq:Fngen}
\boxed{\partial_\tau\langle\ln(1+n)\rangle =\left\langle\frac{1}{2(1+n)}\sum_{a=1}^{\Nf}\frac{\langle\delta f_a\rangle_{\delta\tau}}{\delta\tau}-\frac{1}{8(1+n)^2}\sum_{a,b=1}^{\Nf}\frac{\langle\delta f_a\delta f_b\rangle_{\delta\tau}}{\delta\tau}\right\rangle\,.}
\Eeq
To make further progress, let us first turn to more detailed understanding of $\delta f_a$ and other $\delta\lambda_a$.


\subsection{Increments in \texorpdfstring{$\R$}{R}} 

\begin{figure}[t]
   \centering
   \begin{tikzpicture}
    \node[anchor=south west,inner sep=0] (image) at (0,0) {\includegraphics[width=\textwidth]{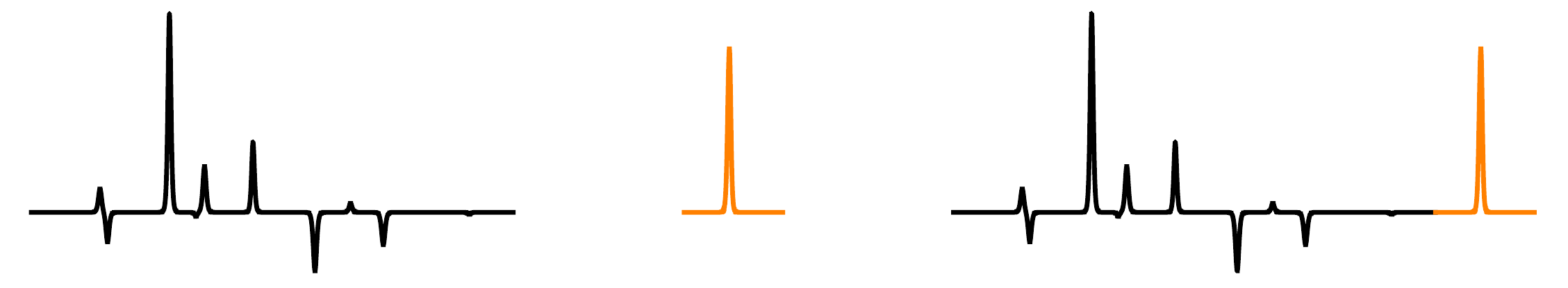}};
    \begin{scope}[x={(image.south east)},y={(image.north west)}]
     \node at (0.38,0.25) {\Large $+$};
     \node at (0.56,0.25) {\Large $=$};
     \node at (0.475,-0.11) {$\M_{j+1}=\I+\m_{j+1}$};
     \node at (0.82,-0.20) {$\begin{aligned}
     						\M (j+1) &= \M_{j+1} \M(j)\\
     						\R (j+1) &= \M_{j+1} \R(j) \M_{j+1}^{\dagger}
					\end{aligned}$};
     \node at (0.18,-0.20) {$\begin{aligned}
     						&\M (j) \\
     						&\R (j) = \M(j)\M^{\dagger}(j)
					\end{aligned}$};
   \end{scope}%
   \end{tikzpicture}%
   \caption{Upon the addition of a scattering with transfer matrix $\M_{j+1}$, the `squared' transfer matrix $\R(j+1)$ is obtained from a quasi-similar transformation of the matrix $\R(j)$. In the weak scattering regime, this amounts to a small change $\delta\R=\R(j+1)-\R(j)$.}
   \label{fig:extrapeak}
\end{figure}
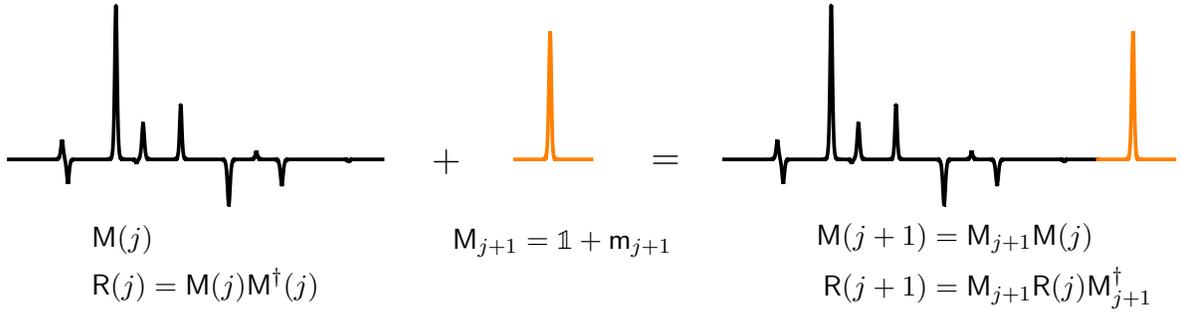%

\noindent Our next task is to find expressions for $\delta \lambda_a$. To this end, let us consider the change in $\R$ due to a single additional scattering (see fig.~\ref{fig:extrapeak}) 
\Beq
\label{eq:dR1}
&\R(j+1)=\R(j)+\delta \R\,,\\
&\delta\R=\R(j)\m^\dagger_{j+1}+h.c.+\m_{j+1}\R(j)\m^\dagger_{j+1}\,.
\Eeq
We will often suppress $``(j)"$ as an argument of the matrices. 

The increment in $\R$, $\delta\R=\delta(\U\F\U^\dagger)$, can be expressed in terms of increments in the eigenvalue matrix $\F$ and the angular matrix $\U$ (up to second order in the increments) as follows:
\Beq
\label{eq:dR2}
\delta \R&=\U\left(\delta \F+[\U^\dagger \delta\U,\F]-[\U^\dagger \delta\U,\F]\U^\dagger \delta\U+[\U^\dagger \delta\U,\delta\F]\right)\U^\dagger\,.
\Eeq
Comparing the expressions for $\delta \R$ in eqns. \eqref{eq:dR1} and \eqref{eq:dR2}, we get
 \Beq
 \label{eq:dFdU}
&\delta \F+[\U^\dagger \delta\U,\F]-[\U^\dagger \delta\U,\F]\U^\dagger \delta\U+[\U^\dagger \delta\U,\delta\F] =\G\,,
\Eeq
where
\Beq
&\G\equiv \left(\Q\F+h.c.\right)+\Q\F\Q^\dagger\,,\qquad \textrm{with}\qquad \Q\equiv\U^\dagger\m_{j+1}\U\,.
\Eeq
Equation \eqref{eq:dFdU} expresses the increments $\delta\F$ and $\delta\U$ in terms of $\G$ which contains information about the additional scattering via $\m_{j+1}$, as well as $\F$ and $\U$. As is evident from the above equation the increments $\delta\F$ and $\delta\U$ will each contain $\F$- and $\U$-related variables. As mentioned earlier, each additional scattering is assumed to make a small difference, hence it is convenient to define the following new quantities with the perturbation order in mind
\Beq
&\delta\F\equiv \delta\F^{(1)}+\delta\F^{(2)}\,,\qquad
&\delta\U\equiv\delta\U^{(1)}+\delta\U^{(2)}\,,\qquad \G^{(1)}\equiv \left(\Q\F+h.c\right)\,,\qquad \G^{(2)}\equiv \Q\F\Q^\dagger\,,\\
\Eeq
which then yields
\Beq
\delta\F^{(1)}+[\U^\dagger \delta\U^{(1)},\F]= \G^{(1)}\,,\\
\delta\F^{(2)}+[\U^\dagger \delta\U^{(2)},\F]-[\U^\dagger \delta\U^{(1)},\F]\U^\dagger \delta\U^{(1)}+[\U^\dagger \delta\U^{(1)},\delta\F^{(1)}] =\G^{(2)}.
\Eeq
Recall that $\F$ and $\U$ are $2\Nf\times2\Nf$ block matrices, which are constructed out of $\Nf\times\Nf$ blocks of functions of $\u$ and $\f$ (and $\tf$) matrices. The above expressions for $\delta\U^{(i)}$ and $\delta\F^{(i)}$ can be unpacked in terms of $\f$ and $\u$ directly to yield 
\begin{subequations}  
\begin{align}
\label{eq:df1}
&\delta\f^{(1)}+[\u^\dagger\delta\u^{(1)},\f]=\g^{(1)}\,,\\ \label{eq:df2}
&\delta\f^{(2)}+[\u^\dagger\delta\u^{(2)},\f]-[\u^\dagger\delta\u^{(1)},\f]\u^\dagger\delta\u^{(1)}+[\u^\dagger\delta\u^{(1)},\delta\f^{(1)}]=\g^{(2)}\,,\\ \label{eq:df3}
&\delta\tilde{\f}^{(1)}+\u^\dagger\delta \u^{(1)} \tilde{\f}-\tilde{\f}\u^T\delta \u^{*(1)}=\tilde{\g}^{(1)}\,,\\ \label{eq:df4}
&\delta\tilde{\f}^{(2)}+\u^\dagger\delta \u^{(2)} \tilde{\f}-\tilde{\f}\u^T\delta \u^{*(2)}-(\u^\dagger\delta \u^{(1)} \tilde{\f}-\tilde{\f}\u^T\delta \u^{*(1)})\u^T\delta\u^{*(1)}+\u^\dagger\delta \u^{(1)} \delta\tilde{\f}^{(1)}-\delta\tilde{\f}^{(1)}\u^T\delta \u^{*(1)}=\tilde{\g}^{(2)}\,,
\end{align}
\end{subequations}
where $\g^{(i)}$ is the top left $\Nf\times \Nf$ block of $\G^{(i)}$, and  $\tilde{\g}^{(i)}$ is the top right $\Nf\times \Nf$ block of $\G^{(i)}$. Although the previous system represents $4\Nf^2$ equations for the $2\Nf(\Nf+1)$ unknowns, there is no issue of overdetermination as, for example, the off-diagonal components of the first and second equations can be checked to be degenerate, leading to a system of $\Nf(\Nf+1)$ equations for the components of $\delta\f^{(1)}$ and $\delta\u^{(1)}$, and similarly for the second-order terms. 

At the level of the variables that parametrize $\f$ and $\u$, the above equations can be solved for each $\delta\lambda_a^{(1)}$ and $\delta\lambda_a^{(2)}$ in terms of the parameters characterizing the scatterers (elements of $\m_j$) as well as $\{\lambda_b\}$. In component form the solution can be written explicitly. The first order correction to $\f$ can be directly obtained from the diagonal elements of eq.~\eqref{eq:df1},
\beq\label{eq:deltaf1}
\delta f^{(1)}_a=g_{aa}^{(1)}\,.
\eeq
For the second-order change in ${\bsf f}$, eq.~\eqref{eq:df1} yields for the off-diagonal components
\beq\label{eq:udu11}
(u^{\dagger}\delta u^{(1)})_{ab} = \frac{g_{ab}^{(1)}}{f_b-f_a}\,,\qquad (a\neq b)\,.
\eeq
In turn, this expression can be substituted into eq.~\eqref{eq:df2} to obtain the diagonal entries, which correspond to
\beq\label{eq:deltaf2}
\delta f_a^{(2)} =  g_{aa}^{(2)} + \sum_{c\neq a} \frac{g_{ac}^{(1)}g_{ca}^{(1)}}{f_a-f_c} \,.
\eeq
The first-order corrections for the entries of $\u$ can be obtained from the off-diagonal components of eq.~\eqref{eq:df1}, and from the diagonal entries of eq.~\eqref{eq:df3}. These lead to the relations eq.~\eqref{eq:udu11} and 
\beq
(u^{\dagger}\delta u^{(1)})_{aa} = \frac{\tilde{g}_{aa}^{(1)}}{2\tilde{f}_a} - \frac{f_a}{2\tilde{f}_a^2}\,\delta\! f_a\,.
\eeq
These two equations form a linear system for the perturbations of $\u$, which can be solved explicitly as 
\beq\label{deltau1}
\delta u_{ab}^{(1)} = \sum_{c\neq b}\frac{u_{ac}g_{cb}^{(1)}}{f_b-f_c} + \frac{u_{ab}}{2\tilde{f}_b^2}\left( \tilde{f}_b \tilde{g}_{bb}^{(1)} -  f_b g_{bb}^{(1)} \right)\,.
\eeq
Analogously, one can solve for the second-order corrections for $\u$ components, which are given by
\begin{align} \notag 
\delta u_{ab}^{(2)} &= \sum_{c\neq b} \frac{u_{ac}}{f_b-f_c}\left[ g^{(2)}_{cb} + g^{(1)}_{cd}(u^{\dagger}\delta u^{(1)} )_{db} - g^{(1)}_{bb}(u^{\dagger}\delta u^{(1)})_{cb} \right]\\ \label{deltau2}
&\quad + \frac{u_{ab}}{2\tilde{f}_b}\left[\tilde{g}^{(2)}_{bb} - \delta\tilde{f}^{(2)}_b - \tilde{g}^{(1)}_{bd}(u^{\dagger}\delta u^{(1)})_{db}^* - \delta\tilde{f}^{(1)}_b(u^{\dagger}\delta u ^{(1)})_{bb} - \tilde{f}_b(\delta u^{(1)\dagger}\delta u^{(1)})_{bb}\right]\,.
\end{align}

To make further progress, we need an explicit form of $\m_j$. Below we will choose Dirac-delta scatterers. 
 Once the form of $\m_j$ is specified, we can solve, in principle, for the increments in the parameters of $\delta\lambda_a^{(i)}$ for arbitrary $\Nf$. In practice this is a non-trivial calculation made particularly onerous by the large number of parameters ($\Nf^2$) in $\u$ and ($\Nf$) in $\f$. 

\subsection{Dirac Delta Scatterers}

The general formalism we have presented is independent of the precise form of the scatterers ($\m_j$), but having a concrete simple example in mind makes things simpler to present and also allows for explicit calculations. With this in mind, as long as we restrict ourselves to wavenumbers $k\ll w^{-1}$, we approximate $m^{\rm s}_{ab}(\tau)$ as follows:
\Beq\label{eq:stomod}
m^{\rm s}_{ab}(\tau)=2\sqrt{\omega_a\omega_b}\sum_{j=1}^{\Ns} \Lambda_{ab}(\tau_j)\delta(\tau-\tau_j)\,,
\Eeq
where $\Ns$ is the number of events, $\tau_j$ are uniformly distributed, and $\delta(\tau-\tau_j)$ are Dirac Delta functions.  For each $\tau_j$, the $\Nf\times\Nf$ elements of $\Lambda_{ab}(\tau_j)$ which characterize the strength of the scatterers are drawn from some distribution. We assume that the distributions are identical and {
\parfillskip=0pt
\parskip=0pt
\par}
\begin{wrapfigure}{R}{0.5\textwidth}
  \centering
   \includegraphics[width=0.45\textwidth]{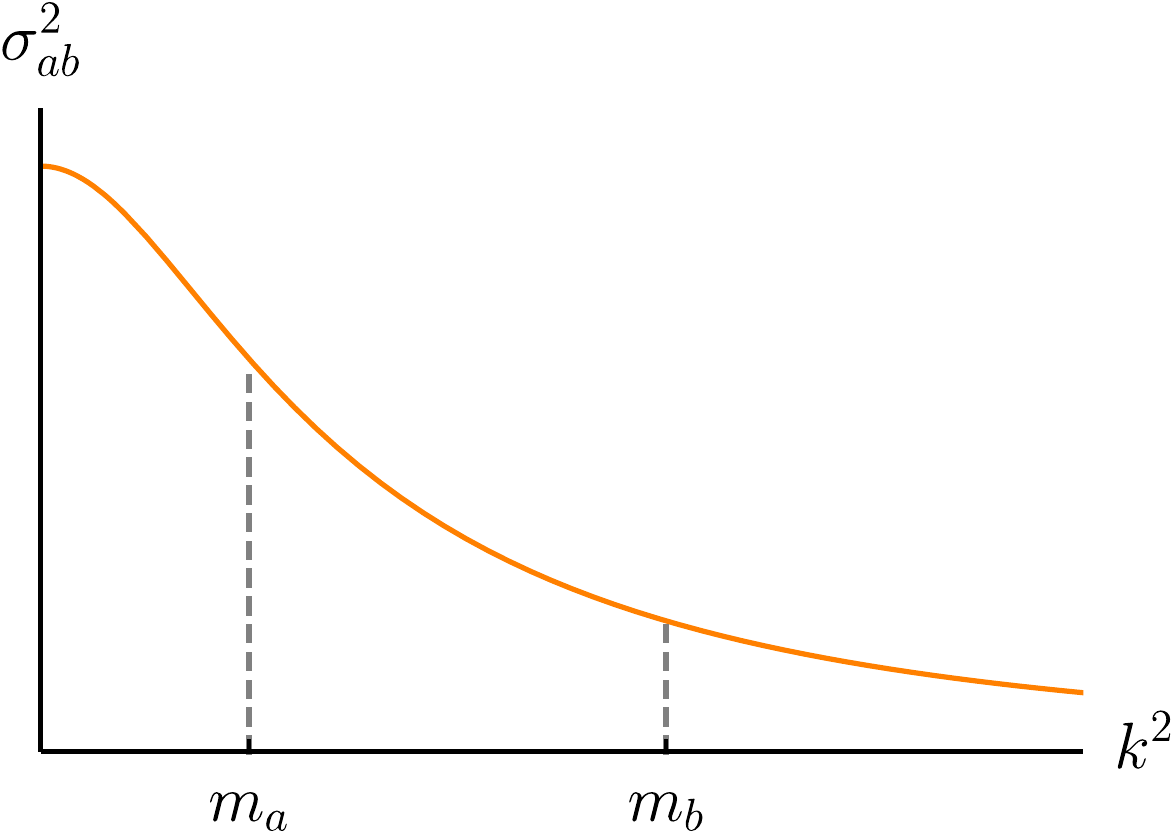}
  \caption{The dependence of the self/cross-coupling variances $\sigma_{ab}^2\propto\left[(k^2+m_a^2)(k^2+m_b^2)\right]^{-1/2}$ on the wavenumber $k$. }\label{fig:sigmap}
\end{wrapfigure}
\noindent independent for all $\tau_j$. Note that the elements of $\Lambda_{ab}(\tau_j)$ must be symmetric with respect to $a$ and $b$. The scaling with $2\sqrt{\omega_a\omega_b}$ is for future convenience. We will assume that for each $\tau_j$
\Beq \label{eq:LambdaSigmaDef}
\begin{aligned}
\langle \Lambda_{ab}(\tau_j)\rangle &= 0\,,\\
\langle \Lambda_{ab}\Lambda_{cd} \rangle &= \sigma^2_{ab}(\delta_{ac}\delta_{bd}+\delta_{ad}\delta_{bc})\,.
\end{aligned}
\Eeq
With the parametrization (\ref{eq:stomod}), the scat\-te\-ring strength variances $\sigma_{ab}^2$ are functions of the wavenumber $k$ and the corresponding scalar field masses (see fig.~\ref{fig:sigmap}).

By imposing appropriate junction conditions on $\chi^a$ at $\tau=\tau_j$, the $2\Nf\times 2\Nf$ transfer matrix at $\M_j$ for the Dirac delta function scatterers evaluates to
\begin{equation}
\label{eq:Delta1}
\M_j = \begin{pmatrix} \p_j^*&0 \\ 0&\p_j\end{pmatrix}\begin{pmatrix} \mathbb{1}+i\mathsf{\Lambda}_j&i\mathsf{\Lambda}_j \\ -i\mathsf{\Lambda}_j&\mathbb{1}-i\mathsf{\Lambda}_j \end{pmatrix}\begin{pmatrix} \p_j&0 \\ 0&\p_j^* \end{pmatrix}=\I+i\underbrace{\begin{pmatrix} \p_j^*&0 \\ 0&\p_j\end{pmatrix}\begin{pmatrix} \mathsf{\Lambda}_j&\mathsf{\Lambda}_j \\ -\mathsf{\Lambda}_j&-\mathsf{\Lambda}_j \end{pmatrix}\begin{pmatrix} \p_j&0 \\ 0&\p_j^* \end{pmatrix}}_{\m_j}\,,
\end{equation}
where
\begin{equation}
\p_j\equiv \begin{pmatrix}
e^{i\omega_1\tau_j} &  0  & \ldots & 0\\
0  &  e^{i\omega_2\tau_j} & \ldots & 0\\
\vdots & \vdots & \ddots & \vdots\\
0  &   0 & \ldots & e^{i\omega_\Nf\tau_j}	
\end{pmatrix}\,,
\qquad \textrm{and}\qquad
\left[\L_j\right]_{ab} = \Lambda_{ab}(\tau_j)\,.
\end{equation}
The right hand sides of eqns.~\eqref{eq:df1}-\eqref{eq:df4} for the delta function scatterers can be written as
\Beq \label{eq:gs}
\g^{(1)}&\equiv [\Gamma,\f] +[\Sigma,\tilde{\f}]  +\tilde{\f} (\Sigma+\Sigma^*)\,,\\
\g^{(2)}&\equiv -\Gamma\f \Gamma+\Gamma\tilde{\f}\Sigma^*-\Sigma \tilde{\f} \Gamma+\Sigma{\f}\Sigma^*\,,\\
\tilde{\g}^{(1)}&\equiv\Gamma \tilde{\f}+\Sigma\f+\f\Sigma-\tilde{\f}\Gamma^*\,,\\
\tilde{\g}^{(2)}&\equiv\Gamma\f \Sigma-\Gamma\tilde{\f}\Gamma^*+\Sigma \tilde{\f} \Sigma-\Sigma{\f}\Gamma^*\,,
\Eeq
where we have defined for the $j+1$-th scatterer
\Beq
\Sigma \equiv i\u^\dagger\p_{j+1}^\dagger \L_{j+1}\p^*_{j+1}\u^*\,,\qquad{\rm and}\qquad
\Gamma \equiv i\u^\dagger\p_{j+1}^\dagger \L_{j+1}\p_{j+1}\u\,.
\Eeq
Note that $\u$ is evaluated at $j$ rather than $j+1$.\\

With the explicit form of the local transfer matrix at hand, we can immediately calculate the disorder-averaged quantities that appear in the expression for the typical occupation number defined in eq.~\eqref{eq:Fngen}, for any number of fields. Namely, for the coefficients that depend on the first-order correction to $\f$, we can rewrite eq.~\eqref{eq:deltaf1} in terms of the matrix $\g^{(1)}$ given in eq.~\eqref{eq:gs}. The following explicit expression for the perturbation $\delta f_a^{(1)}$ is obtained:
\begin{align}\notag
\delta f_a^{(1)} &= i \tilde{f}_a  \sum_{b,c,d,e} \left(u_{ab}^{\dagger}p_{bc}^{\dagger}\Lambda_{cd}p_{de}^*u_{ea}^*  - u_{ab}^{T}p_{bc}^{T}\Lambda_{cd}p_{de} u_{ea} \right)\\
&= i \tilde{f}_a \sum_{b,c} \left(p_b^*p_c^* u_{ab}^{\dagger}\Lambda_{bc}u_{ac}^{\dagger}  - p_b p_c u_{ba} \Lambda_{bc}  u_{ca} \right)\,.
\end{align}
In the second line we have made use of the fact that ${\p}$ is diagonal. Consider now the expectation value $\langle \delta f_a^{(1)}\delta f_b^{(1)}\rangle_{\delta\tau}$. This product contains phase factors due to the presence of the components of $\p$, which can be explicitly evaluated, as each $\tau_j$ is assumed to be uniformly distributed in the interval $[(j-1)\delta\tau,j\delta\tau]$. For example,
\beq
\label{eq:phaseavg}
\frac{\langle p_a p_b p^{*}_c p^*_d\rangle_{\delta\tau}}{\delta\tau} = \frac{\langle e^{i \Delta\omega\tau_j}\rangle_{\delta\tau}}{\delta\tau} = \frac{\sin(\Delta\omega\delta\tau/2)}{\Delta\omega\delta\tau/2}\,e^{i\Delta\omega(j-1/2)\delta\tau}\,.
\eeq
where $\Delta\omega=\omega_a+\omega_b-\omega_c-\omega_d$. In what follows, and throughout this paper, we will implicitly assume that the time period of oscillations is much smaller than the mean free path determined by the separation between events, $\omega\delta\tau\gg1$; this is consistent with the treatment of the dynamics as free in-between scatterings. This assumption implies that, within the product $\langle \delta f_a^{(1)}\delta f_b^{(1)}\rangle_{\delta\tau}$, we can disregard terms that contain unpaired ${\bsf p}$ and ${\bsf p}^*$, as they will vanish after averaging over scatterers; additionally $p^*_a p_b\rightarrow \delta_{ab}$ upon averaging. Recalling the statistical properties of $\Lambda_{ab}$ in eq.~\eqref{eq:LambdaSigmaDef}, we can then write\footnote{In eq.~(\ref{varf1}) (and in what follows) we have denoted $\langle \delta f_a^{(1)}\delta f_b^{(1)} \rangle_{\delta\tau} \equiv \lim_{\delta\tau\rightarrow 0} \langle \delta f_a^{(1)}\delta f_b^{(1)} \rangle_{\delta\tau}/\delta\tau$. Whenever there is no potential for confusion we will also suppress the $\delta\tau$ subindex, to avoid cumbersome expressions.}
\beq\label{varf1}
\langle \delta f_a^{(1)}\delta f_b^{(1)} \rangle_{\delta\tau} =  2\tilde{f}_a\tilde{f}_b \sum_{c,d} \sigma_{cd}^2  \left( u_{ac}^{\dagger} u_{ad}^{\dagger} u_{cb}u_{db} + u_{bc}^{\dagger} u_{bd}^{\dagger} u_{ca}u_{da} \right)\,.
\eeq

The second-order change in ${\bsf f}$ due to a scattering can be computed in an analogous way, directly upon averaging (\ref{eq:deltaf2}) over scatterings, a procedure which leads to the following expression for the mean second-order perturbation,
\beq\label{varf2}
\langle \delta f_a^{(2)} \rangle_{\delta\tau} = \left\langle g_{aa}^{(2)} + \sum_{b\neq a} \frac{g_{ab}^{(1)}g_{ba}^{(1)}}{f_a-f_b} \right\rangle_{\delta\tau}\,,
\eeq
where
\beq
\langle g_{aa}^{(2)}\rangle_{\delta\tau} = \sum_{b,c,d} f_d\, \sigma_{bc}^2 \left[(1+\delta_{bc})u^{\dagger}_{ab} u_{bd} u^{\dagger}_{dc} u_{ca} + 2 u^{\dagger}_{ab} u_{ba} u^{\dagger}_{dc} u_{cd}\right]\,,
\eeq
and
\beq
\begin{aligned}
\left\langle \sum_{b\neq a} \frac{g_{ab}^{(1)}g_{ba}^{(1)}}{f_a-f_b} \right\rangle_{\delta\tau} &= \sum_{\substack{b,c,d\\ b\neq a}} \sigma_{cd}^2 \left[(f_a-f_b) \left(u^{\dagger}_{ac} u_{ca} u^{\dagger}_{bd} u_{db} + \delta_{cd} u^{\dagger}_{ac}u_{db} u^{\dagger}_{bc} u_{da}\right)\right.\\
&\qquad\qquad\ \ + \frac{\tilde{f}_b^2 + \tilde{f}_a^2}{f_a-f_b}  \left. \left(u^{\dagger}_{ac} u_{ca} u^{\dagger}_{bd} u_{db} +  u^{\dagger}_{ac}u_{cb} u^{\dagger}_{bd} u_{da}\right) \right]\,.
\end{aligned}
\eeq
Substitution of (\ref{varf1}) and (\ref{varf2}) into (\ref{eq:Fngen}) then yields
\begin{align}\notag
\partial_{\tau}\langle \ln(1+n)\rangle  &= \sum_{a,b,c,d} \Bigg\langle \frac{\sigma_{cd}^2}{2(1+n)}\Bigg[f_a\,\big(2u_{ac}^{\dagger}u_{bd}^{\dagger}u_{db}u_{ca} + u_{ac}^{\dagger}u_{bd}^{\dagger}u_{da}u_{cb} + \delta_{cd}u_{ac}^{\dagger}u_{bc}^{\dagger}u_{cb}u_{ca}\big)\\ \notag
&\qquad\qquad\qquad -\frac{\tilde{f}_a\tilde{f}_b}{(1+n)}    u_{ac}^{\dagger} u_{ad}^{\dagger} u_{cb}u_{db}  \Bigg] \Bigg\rangle \\ \label{eq:dtnfg}
& \ \ + \sum_{\substack{a,b,c,d\\ c\neq a}} \Bigg\langle \frac{\sigma_{cd}^2}{2(1+n)}\frac{\tilde{f}_c^2 + \tilde{f}_a^2}{f_a-f_c}   \left(u^{\dagger}_{ab} u^{\dagger}_{cd} u_{ba}  u_{dc} +  u^{\dagger}_{ab} u^{\dagger}_{cd} u_{bc}  u_{da}\right) \Bigg\rangle \,.
\end{align}
This is an explicit expression for the evolution of the typical occupation number for arbitrary $\Nf$ and $\sigma_{ab}$ (assuming Dirac-delta function scatterers).
However, note that the angular brackets denote integration with respect to probability distribution $dP(\{\f,\u\})$. In the absence of an explicit parametrization for $\u$ and the solution to the FP equation, equation~(\ref{eq:dtnfg}) is of limited applicability. In the next section we will evaluate it under the assumption of a $\u$-flat probability distribution and for a statistically isotropic $\L$ (i.e. $\sigma^2_{ab}=\sigma^2$). In the subsequent sections, we will evaluate this expression in general for the $\Nf=1,2$ cases, and for $\Nf>2$ under some restrictive conditions for the scattering strengths $\sigma_{ab}^2$.

\section{The Maximum Entropy Approximation}\label{sect:MEA}

In the previous section, we have laid the foundations necessary to compute the solution of the FP equation that determines the instantaneous dependence of the probability distribution $P$ with respect to the parameters in $\R$. However, before proceeding to such analysis, we will review in this section the predictions of the Maximum Entropy approximation, and state our expectations for when we eventually compare them against the analytical result obtained from the integration of the FP equation.

The probability density $P$ associated with the MEA corresponds to that which maximizes the (Shannon) entropy functional \cite{mello2004quantum},
\beq\label{eq:Sdef}
S[P] = -\langle \ln P\rangle_{\delta\tau}\,,
\eeq
subject to the constraints that the local mean particle production rate is known, and that the evolution of the transfer matrix $\M$ is under perturbative control, $\M_{\tau+\delta\tau}\rightarrow\M_{\tau}$ as $\delta\tau\rightarrow 0$. With this definition, the MEA recipe is said to provide the least biased estimate of the $P$ consistent with the (fixed) local production rate. Under these approximations, it can be shown that the probability density is independent of $\u$, $P(\{\f,\u\})=P(\{\f\})$. Moreover, for the typical occupation number $n_{\rm typ}$, the MEA implies the late-time result \cite{Amin:2015ftc}
%
\beq\label{eq:meaa}
\partial_{\tau}\langle \ln(1+n)\rangle_{\rm MEA} \longrightarrow \frac{2\Nf}{\Nf+1}\,\mu_j \,,
\eeq
where $\mu_j$ denotes the instantaneous mean particle production rate,
\beq
\mu_j \equiv \frac{1}{\Nf}\,\frac{\langle n_j\rangle_{\delta\tau}}{\delta\tau}\,.
\eeq
Here $\langle n_j\rangle_{\delta\tau}$ corresponds to the {\em local} change in the occupation number due to a single scattering event. For Dirac-delta scatterers, it can be easily evaluated in terms of the scattering strengths as
\beq\label{eq:meab}
\frac{\langle n_j\rangle_{\delta\tau}}{\delta\tau} \;=\; \frac{1}{4}\left\langle \Tr\left[\M_j \M^\dagger_j -\I\right] \right\rangle_{\delta\tau} \;=\; \left\langle \Tr\,\L_j^2 \right\rangle_{\delta\tau} \;=\; \sum_{a,b=1}^{\Nf}\sigma_{ab}^2(1+\delta_{ab})\,.
\eeq
where in the last equality we have used the defining relation (\ref{eq:LambdaSigmaDef}). Thus, the MEA-based result is 
\Beq
\label{eq:meac}
\partial_{\tau}\langle \ln(1+n)\rangle_{\rm MEA} \longrightarrow \frac{2}{\Nf+1}\,\left\langle \Tr\,\L_j^2 \right\rangle_{\delta\tau} \,.
\Eeq
\\
\noindent{\bf Factorizability, the Haar measure \& the MEA}: The independence of $P$ on $\u$ within MEA implies that the differential probability may be written as
\beq\label{eq:dPhaar}
dP(\{\f,\u\})= P(\{\f\})\,d\mu(\u)\,,
\eeq
where $d\mu(\u)$ denotes the invariant (Haar) measure on the group of unitary matrices $U(\Nf)$. The previous expression is a key prediction of the MEA approximation, and, as we now show, it is the single result responsible for the trace formula in eq.~\eqref{eq:meab}. For these purposes, let us now assume that the probability density satisfies eq.~(\ref{eq:dPhaar}), {\em without} assuming that it extremizes the Shannon entropy in eq.~(\ref{eq:Sdef}). As it was shown in the previous section, in the case of Dirac delta scatterers, the general expression for time-evolution of the function $F(n)=\ln(1+n)$ can be written as eq.~(\ref{eq:dtnfg}) in terms of an average with respect to the measure $dP(\{\f,\u\})$. Under the assumption (\ref{eq:dPhaar}), the averages for the monomial functions of the components of $\u$ and $\u^{\dagger}$ can be immediately computed in terms of the Weingarten relations \cite{mello2004quantum}, which for monomials of ranks 2 and 4 correspond to
\beq
\int u_{ab}u_{a'b'}^* \, d\mu(\u) \;=\; \frac{1}{\Nf}\delta_{aa'}\delta_{bb'}\,,
\eeq
and
\begin{align}\notag
\int u_{a_1b_1}u_{a_2b_2}u_{a_1'b_1'}^*u_{a_2'b_2'}^*\, d\mu(\u) \;&=\; \frac{\delta_{a_1a_1'}\delta_{a_2a_2'}\delta_{b_1b_1'}\delta_{b_2b_2'} + \delta_{a_1a_2'}\delta_{a_2a_1'}\delta_{b_1b_2'}\delta_{b_2b_1'} }{\Nf^2-1}\\
&\quad - \frac{\delta_{a_1a_1'}\delta_{a_2a_2'}\delta_{b_1b_2'}\delta_{b_2b_1'} + \delta_{a_1a_2'}\delta_{a_2a_1'}\delta_{b_1b_1'}\delta_{b_2b_2'} }{\Nf(\Nf^2-1)}\,.
\end{align}
By making use of these relations, the general expression (\ref{eq:dtnfg}) reduces to
\beq
\partial_{\tau}\langle \ln(1+n)\rangle_{\rm Haar} \;=\; \left\langle \frac{1}{\Nf(1+n)}\sum_{a=1}^{\Nf} f_a - \frac{1}{2\Nf(\Nf+1)(1+n)^2} \sum_{a=1}^{\Nf} \tilde{f}_a^2 \right\rangle \,\sum_{b,c=1}^{\Nf}\sigma_{bc}^2(1+\delta_{bc})\,.
\eeq

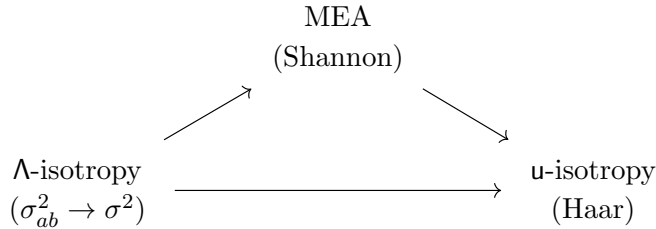
\begin{figure}[t] 
   \centering
   \begin{tikzcd}
{} & \begin{tabular}{c}
MEA \\
(Shannon)
\end{tabular} \arrow{dr}{} \\
\begin{tabular}{c}
$\L$-isotropy \\
($\sigma^2_{ab}\rightarrow\sigma^2$)
\end{tabular} \arrow[ur,rightarrow]{} \arrow{rr}{} && \begin{tabular}{c}
$\u$-isotropy \\
(Haar)
\end{tabular} 
\end{tikzcd}
\caption{The MEA relies on the extremization of the Shannon entropy for the probability in the fundamental strip; it implies a flat (Haar) distribution over the $U(\Nf)$ group parametrized by $\u$. We argue that the MEA is an exact solution of the FP equation in the $\L$-isotropic limit, for which a $\u$-flat distribution is also obtained.}
   \label{fig:triangle}
\end{figure}

In the right-hand side we recognize the trace in eq.~(\ref{eq:meab}). The prefactor is the expectation value of a complicated function of the eigenvalues $f_a$. Nevertheless, in the late time limit, it is expected that the occupation number $n\gg1$. Moreover, in the asymptotic regime, it is expected that the eigenvalues of $\M$, itself a product of a large number of transfer matrices, separate, and the asymptotic growth is purely exponential, implying that we can assume that a single eigenvalue dominates the evolution at late times~\cite{crisanti1993products,oseledec1968multiplicative}. Recalling the relation (\ref{eq:njfa}) we can therefore write
\beq\label{eq:dthaar}
 \partial_{\tau}\langle \ln(1+n)\rangle_{\rm Haar} \;\longrightarrow\; \frac{2 }{\Nf+1} \left\langle \Tr\,\L_j^2 \right\rangle_{\delta\tau}\,.
\eeq
which is none other than the MEA result (\ref{eq:meac}). This result will be of use in the coming sections.\\

\noindent{\bf $\L$-isotropy and the MEA}: The previous calculation demonstrates that factorizability and $\u$-isotropy (see eq.~\eqref{eq:dPhaar}) is sufficient to reproduce the MEA prediction for the evolution of the typical occupation number (a fact already discussed in~\cite{Amin:2015ftc}). However, it tells us nothing about the probability density $P(\{\f\})$. On one hand, within MEA, its form would be determined by the requirement that it extremizes the entropy functional (\ref{eq:Sdef}) and then solving the resulting FP equation for $P(\{\f\})$. On the other hand, in terms of the microscopic model built in section~\ref{sec:TMA}, $P(\{\f\})$ would be determined by the particular set of scattering strengths $\sigma_{ab}$ for which the solution of the FP equation (\ref{eq:FPlambda}) is $\f$-$\u$ factorizable and compatible with (\ref{eq:dPhaar}). It is a-priori unclear under what conditions on the microscopic model that one recovers the MEA-based $P(\{\f\})$.

It has been conjectured that the Maximum Entropy approximation is equivalent to the assumption that the interacting fields are statistically equivalent and maximally mixed with a random but uniform distribution of non-adiabatic events~\cite{2012PhRvB..86a4205X,Amin:2015ftc}. In the context of the present study, it would correspond to   $\sigma_{ab}^2=\sigma^2$ (i.e. a statistically isotropic $\L$). Using our general expression in eq.~\eqref{eq:dtnfg}, we can analytically  verify this conjecture in the context of the typical occupation number. Using $\sigma_{ab}^2=\sigma^2$  in eq.~\eqref{eq:dtnfg}, we immediately get 
\begin{align}\notag
\partial_{\tau}\langle \ln(1+n)\rangle_{\L\text{-iso}} \;&=\; \left\langle \frac{1}{2(1+n)}\left(2(\Nf+1)\sum_{a=1}^{\Nf} f_a - \frac{1}{(1+n)} \sum_{a=1}^{\Nf} \tilde{f}_a^2\right) \right\rangle \,\sigma^2\,,\\ 
&\rightarrow\; 2\Nf\,\sigma^2\,,\notag\\
&=\frac{2}{\Nf+1}\langle\Tr\, \L^2_{j}\rangle_{\delta\tau}\bigg|_{\sigma_{ab}^2=\sigma^2}
\,,\label{eq:isores}
\end{align}
where the second line is valid in the late-time limit under the assumption of a dominant eigenvalue. That is, $n=(1/2)\sum_{a}^{\Nf}(f_a-1)\rightarrow f_{\rm max}/2$. In the third line we have used a straightforward substitution of $\sigma_{ab}^2=\sigma^2$ in eqns.~(\ref{eq:meab}). We have shown that the result in eqns.~(\ref{eq:isores}) corresponds to the $\L$-isotropic limit of the MEA expression (\ref{eq:meac}). In the upcoming sections, we will also realize that $\L$-isotropy $\rightarrow \u$-isotropy (in a somewhat restricted sense relevant for the typical occupation number). Fig.~\eqref{fig:triangle} is a graphical illustration of which assumption leads to what conclusion.

While we have proven that in the $\L$-isotropic case, the MEA result for the occupation number matches the result from explicit calculation, we have yet to see how the exact results deviate (or even if they deviate) from MEA based results when $\L$-isotropy is not assumed. In the following sections, we will carry the full analysis assuming neither MEA nor $\u$ nor  $\L$-isotropy for the $\Nf=1,2$ cases, and in a restricted sense, for the general $\Nf>2$ case. 


%


\section{A Warm-up Example: the \texorpdfstring{$\Nf=1$}{Nf=1} Case}\label{sect:onefield}

Let us consider first the simplest case, that of a single field. The unitary matrix $\u$ reduces simply to a phase, $u=e^{i\phi}$, meaning that the $\R$ matrix contains only the two real variables $\{\lambda_a\}=\{f,\phi\}$. Straightforward substitution into eqns.~(\ref{eq:df1})-(\ref{eq:df4}) leads to the following expressions for $\{\delta\lambda_a\}$,
\begin{subequations}  
\begin{align}
\delta f^{(1)} &= g^{(1)}\,,\\
\delta f^{(2)} &= g^{(2)}\,, \\
\delta \phi^{(1)}  &= -\frac{i}{2\tilde{f}^2}\left( \tilde{f}\tilde{g}^{(1)} - f g^{(1)} \right)\,,\\
\delta \phi^{(2)} & = -\frac{i}{2\tilde{f}}\left[\tilde{g}^{(2)}-\frac{f}{\tilde{f}}\,g^{(2)} - \frac{(\tilde{g}^{(1)})^2}{2\tilde{f}} + \frac{(f^2+1)(g^{(1)})^2}{2\tilde{f}^3} \right]\,,
\end{align}
\end{subequations}
where for a single delta function scatterer we have (see eq.~\eqref{eq:gs})
\begin{subequations} 
\begin{align}
g^{(1)} & =  i\tilde{f}\Lambda_j \left(e^{-2 i (\phi+\omega\tau_j)}-e^{2 i (\phi+\omega\tau_j)} \right)\,,\\
\tilde{g}^{(1)} &= 2i\Lambda_j\left(\tilde{f} + fe^{-2 i (\phi+\omega\tau_j)} \right)\,,\\
g^{(2)} & =  2f\Lambda_j^2 + \tilde{f}\Lambda_j^2 \left(e^{-2 i (\phi+\omega\tau_j)} + e^{2 i (\phi+\omega\tau_j)} \right)\,,\\
\tilde{g}^{(2)} &= -\tilde{f}\Lambda_j^2\left(1+e^{-4 i (\phi+\omega\tau_j)}\right) - 2f\Lambda_j^2e^{-2 i (\phi+\omega\tau_j)}\,.
\end{align}
\end{subequations}
With the results at hand, we can calculate the coefficients of the FP equation by taking averages over the properties of the scatterer as well as the location of the scatterers. For this purpose we will denote the ($\tau_j$-independent) variance of $\Lambda_j$ by
\beq
\langle \Lambda_j^2\rangle_{\delta\tau} = \sigma^2\,.
\eeq
As it was discussed around eq.~\eqref{eq:phaseavg}, under the assumption $\omega\delta\tau\gg1$, all time-dependent phase factors average to zero. We then find that
\begin{subequations}  
\begin{align}
\langle \delta f^{(1)} \delta f^{(1)}\rangle  &=2\tilde{f}^2\sigma^2\,,\\
\langle \delta f^{(1)} \delta \phi^{(1)}\rangle &= 0\,, \\
\langle \delta \phi^{(1)} \delta \phi^{(1)}\rangle  &= \frac{\sigma^2}{2\tilde{f}^2}(2\tilde{f}^2+f^2)\,,\\
\langle \delta f^{(2)}\rangle &= 2f\sigma^2\,,\\
\langle \delta \phi^{(2)} \rangle  & = 0\,,
\end{align}
\end{subequations}
where we have lost the $\delta\tau$ sub-index for notational simplicity. It is worth noting that all expectation values are independent of the angular variable $\phi$. Thus, let us consider the marginal probability distribution $\int d\phi\, P(f,\phi;\tau)$, which somewhat abusing notation we will continue denoting by $P$. The FP equation~\eqref{eq:FPlambda} satisfied by $P$ becomes
\beq\label{eq:FPsf}
\frac{1}{\sigma^2}\frac{\partial}{\partial\tau}P(f;\tau)=  - 2  \, \frac{\partial}{\partial f}\Big[f\, P(f;\tau)\Big] +   \frac{\partial^2}{\partial f^2} \Big[ \tilde{f}^2 P(f;\tau)\Big] = \frac{\partial}{\partial f}\left[(f^2-1)\frac{\partial}{\partial f}P(f;\tau)\right]\,.
\eeq
Although the exact solution of eq.~\eqref{eq:FPsf} can be found in integral form~\cite{Abrikosov1981997}, we are interested in its behavior in the asymptotic regime $\tau\rightarrow\infty $; in this limit we expect that $n\gg 1$ and hence $f\gg 1$. Therefore, in the right-hand side we may replace $(f^2-1)\rightarrow f^2$. Further switching the independent variable from $f$ to $n$, we find the FP equation
\begin{flalign}
& \text{(late-time)} & \Cen{3}{\frac{1}{\sigma^2}\frac{\partial}{\partial\tau}P(n;\tau) = \frac{\partial}{\partial n}\left[n^2\frac{\partial}{\partial n}P(n;\tau)\right]\,,}      &&  
\end{flalign}
which for an initial condition $P(n;0)=\delta(n)$ has the log-normal solution,
\begin{flalign}\label{eq:FPsol}
& \text{(late-time)} & \Cen{3}{P(n;\tau)\,dn = \frac{1}{\sqrt{4\pi\sigma^2\tau}}\exp\left[-\frac{(\ln n - \sigma^2\tau)^2}{4\sigma^2\tau}\right]\,d\ln n\,.}      &&  
\end{flalign}
Equation (\ref{eq:FPsol}) recovers the single field result derived in~\cite{Amin:2015ftc}; where it was shown that the assumption of maximum entropy implies the independence of the probability density on the angular variable $\phi$, which inevitably leads to the solution in eq.~(\ref{eq:FPsol}) after assuming that the local mean particle production rate is known. 

Having an explicit expression for the probability density, one can in principle calculate all moments of the distribution. For the typical occupation number in eq.~(\ref{eq:ntypd}), we have 
\beq\label{eq:ntyp1}
n_{\rm typ} = e^{\sigma^2 \tau}-1\,.
\eeq

Although the simplicity of the single field case allowed us to calculate a closed form for the probability distribution $P$, the complexity of the analysis increases many-fold even upon the addition of a single additional field, and we will not obtain the full form for $P$ even in the two-field scenario. Nevertheless, as we argued in section~\ref{sect:MEA}, the full solution of the FP equation is not needed to obtain an expression for the (total) occupation number. Namely, if we consider the function of the occupation number $F(n)=\ln(1+n)$, we can immediately write the late time behavior of eq.~(\ref{eq:Fngen}) as
\Beq 
\lim_{\tau\rightarrow\infty} \partial_\tau \langle \ln(1+n)\rangle = \lim_{\tau\rightarrow\infty} \left \langle\frac{1}{1+f}\cdot 2f\sigma^2- \frac{1}{2(1+f)^2}\cdot 2\tilde{f}^2\sigma^2 \right\rangle = \sigma^2\,,
\Eeq
which trivially implies (\ref{eq:ntyp1}). We will continue to exploit the relation in eq.~(\ref{eq:Fngen}) in the two- and multi-field cases discussed below.

To summarize, without considering the MEA, we arrive at a result identical to the one obtained by using the MEA. The equivalence to the MEA arises from the fact the disorder averaged correlators  are independent of $\u$ (parametrized by $\phi$ here); the $\Nf=1$ is special is this sense. Let us now consider the more interesting and more involved case with $\Nf=2$.
\section{Beyond MEA: The \texorpdfstring{$\Nf=2$}{Nf=2} Case}\label{sect:twofield}

For the two field case ($\Nf=2$), the unitary matrix $\u$ can be parametrized in terms of the $U(2)$ Euler angles as follows,
\Beq
\u(\phi,\theta,\psi,\varphi) = 
e^{-\frac{i}{2}\phi}\begin{bmatrix} 
\vspace{0.1cm}
\cos\frac{\theta}{2} \, e^{-\frac{i}{2}(\varphi+\psi)} & -\sin\frac{\theta}{2} \, e^{-\frac{i}{2}(\varphi-\psi)}\\
\sin\frac{\theta}{2}\,e^{\frac{i}{2}(\varphi-\psi)}  & \cos\frac{\theta}{2} \, e^{\frac{i}{2}(\varphi+\psi)}
\end{bmatrix},
\Eeq
where $\phi, \varphi \in [0, \, 2 \pi)$, $\theta \in [0, \, \pi)$, and $\psi \in [0, \, 4 \pi)$~\cite{biedenharn1984angular}. The $\f$ matrix is diag$\{f_1,f_2\}$. Therefore the $\R$ matrix contains six real variables,
\Beq
\{\lambda_a\}= \left\{f_1, \, f_2, \, \phi, \, \theta, \, \psi, \, \varphi \right\}.
\Eeq

\subsection{Calculation of the Correlators}

In complete analogy with the single field case, and after a significant amount of algebra, we find the following expressions for $\{\delta\lambda_a\}$. First, at leading order (see eqns.~(\ref{eq:df1})-(\ref{eq:df4}))
\begin{subequations}  
\begin{align}
\delta{f}_\varrho^{(1)} & \, = \g^{(1)}_{\varrho,\varrho} ,\quad  \varrho \in \{1,2\}, \\ \displaybreak[0]
\delta{\theta}^{(1)}  & \, =\frac{2}{\Delta{f}} \Re{\left(\g^{(1)}_{2,1} e^{i \psi} \right)},  \\ \displaybreak[0]
\delta{\varphi}^{(1)} & \,  = \frac{2}{\Delta{f}}\Im{\left(\g^{(1)}_{2,1}e^{i \psi}\right)} \csc{\theta}\,,
\end{align}
\end{subequations}
where $\Delta f=f_1-f_2$. At next to leading order, we have
\begin{subequations}                    
\begin{align}
\delta{f}_\varrho^{(2)} &\,=\g^{(2)}_{\varrho,\varrho} -(-1)^{\varrho} \frac{|\g^{(1)}_{2,1}|^2} {\Delta{f}}  ,\quad\varrho \in \{1,2\},  \\
\delta{\theta}^{(2)} &\, = \frac{2}{\Delta{f}} \Re{\left( \g^{(2)}_{2,1} e^{i \psi} \right)} +  \frac{1}{\Delta{f}}\left( \delta{f}_2^{(1)}-\delta{f}_1^{(1)} \right)\delta{\theta}^{(1)}+\frac{1}{4}\sin{2\theta}\left( \delta{\varphi}^{(1)} \right)^{2},
\end{align}
\end{subequations}
The leading order coefficients should be subsituted in the expressions above to get the second order contributions explicitly. Note that we have only listed a select number of $\delta\lambda_a$. The rest are provided in Appendix \ref{ap:nf2}.

For evaluating the $\g^{(i)}$ and $\tilde{\g}^{(i)}$ matrices we need the transfer matrix for a single delta function scatterer with $\Nf=2$. This is given by eq.~\eqref{eq:Delta1} with
\Beq
\p_j=\begin{pmatrix}
e^{i\omega_1\tau_j} & 0\\
0  &  e^{i\omega_2\tau_j}	
\end{pmatrix}\,,\qquad \textrm{and}\qquad 
\L_j=\begin{pmatrix}
\Lambda_{11}(\tau_j)& \Lambda_{12}(\tau_j)\\
\Lambda_{21}(\tau_j)  &  \Lambda_{22}(\tau_j)	
\end{pmatrix}\,.
\Eeq
Note that $\Lambda_{12}(\tau_j)=\Lambda_{21}(\tau_j)$. In order to calculate the coefficients of the FP equation we will need the variances of the entries of $\L_j$. These $j$-independent variances will be denoted as follows,
\Beq
\langle (\Lambda_{11})^2\rangle_{\delta\tau} = \sigma^2_{1}\,,\qquad\langle (\Lambda_{22})^2\rangle_{\delta\tau} = \sigma^2_{2}\,,\qquad\langle (\Lambda_{12})^2\rangle_{\delta\tau}=\langle (\Lambda_{12})^2\rangle_{\delta\tau}=\sigma_{\perp}^2\,.
\Eeq
Care should be taken regarding the notation $\sigma_i^2$ and $\sigma_{ab}^2$. In particular 
\Beq
\sigma_{11}^2=\sigma_1^2/2\,,\qquad\sigma_{22}^2=\sigma_2^2/2\,,\qquad\sigma_{12}^2=\sigma_{21}^2=\sigma_{\perp}^2\,.
\Eeq
We apologize for what might seem like confusing notation here; it is a direct consequence of the way we defined $\sigma^2_{ab}$ in eq.~\eqref{eq:LambdaSigmaDef}.\footnote{This notation will also be useful for $\Nf>2$ case.} Moreover, we assume that $\langle e^{i\omega_a\tau_j}\rangle_{\delta\tau} =0$ for $a=1,2$ as well as $\langle e^{i(n\omega_1-m\omega_2)\tau_j}\rangle_{\delta\tau} =0$ as long as $n\omega_1\ne m\omega_2$ with $n,m\in$ Integers.\footnote{We have found that our numerical results match the analytical predictions even when $\omega_1=\omega_2$. In addition, numerical evidence in the multifield case also indicates that the equal $\omega_a$ case is not special.} 

With these assumptions at hand, the computation of the coefficients of the FP equation proceeds in a straightforward way, albeit requiring a significant amount of algebra. The full list of correlators is provided in Appendix~\ref{ap:nf2}. Here, we simply list the ones that are directly needed to complete the calculation:
\begin{align}
\langle \delta f_1^{(1)} \delta f_1^{(1)} \rangle
&\;= \; \tilde{f}_1^2 \, \gamma_1(\theta), \\
\langle \delta f_1^{(1)} \delta f_2^{(1)} \rangle 
&\;= \; 2 \tilde{f}_1 \tilde{f}_2 \cos(2 \psi) \,\gamma_3(\theta), \\
\langle \delta f_2^{(1)}\delta f_2^{(1)}  \rangle &\;= \; \tilde{f}_2^2 \, \gamma_2(\theta),\\
\langle\delta{f}_{1}^{(2)}\rangle &\;= \; \frac{1}{\Delta{f}} \left[ \tilde{f}_1^2 l_1(\theta) -(f_1 f_2 -1) l_3(\theta)  \right], \\
\langle\delta{f}_{2}^{(2)}\rangle &\;= \; -\frac{1}{\Delta{f}} \left[ \tilde{f}_2^2 l_2(\theta) -(f_1 f_2 -1) l_4(\theta)  \right],\\
\langle\delta \theta^{(1)} \delta \theta^{(1)}\rangle &\;=\; 2 \sigma_\perp^2 + \frac{1}{\Delta{f}^2} \left( \tilde{f}_1^2 + \tilde{f}_2^2 + 2\tilde{f}_1 \tilde{f}_2 \cos(2 \psi)  \right) \gamma_6(\theta), \\
\langle\delta{\theta}^{(2)}\rangle &\;= \;-\frac{(f_1+f_2)}{\Delta{f}} \left(\sigma _1^2-\sigma _2^2\right)  \sin{\theta}  -\frac{1}{2} \sin (2 \theta ) \left(\sigma _1^2+\sigma _2^2 -3\sigma_\perp^2 \right)\,.
\end{align}
where the $\gamma$- and $l$-functions, which depend only on the polar angle $\theta$ and the scattering strengths $\sigma_{1,2,\perp}^2$, are defined in the Appendix~\ref{ap:nf2} in eqns.~(\ref{eq:1stgamma})-(\ref{eq:lastl}). We list $\gamma_1(\theta)$ and $l_1(\theta)$ here to provide an idea of their structure.
\Beq
\gamma_1(\theta) & \, =  2 \left[ 
  \sigma_1^2 \cos^4\left( \frac{\theta}{2} \right)
+ \sigma_2^2 \sin^4\left( \frac{\theta}{2} \right)
+ 4 \sigma_\perp^2 \sin^2\left( \frac{\theta}{2} \right) \cos^2\left( \frac{\theta}{2} \right)
\right], \\
l_1(\theta) & \, =  2 \left[ \cos^2 \left(\frac{\theta}{2}\right) \sigma_1^2 + \sin^2 \left(\frac{\theta }{2}\right) \sigma_2^2 + \sigma_\perp^2 \right]\,.
\Eeq

\subsection{The Typical Occupation Number}\label{sec:typical2}

\begin{figure}[t!]
\centering
    \subfloat{\includegraphics[width=0.49\textwidth]{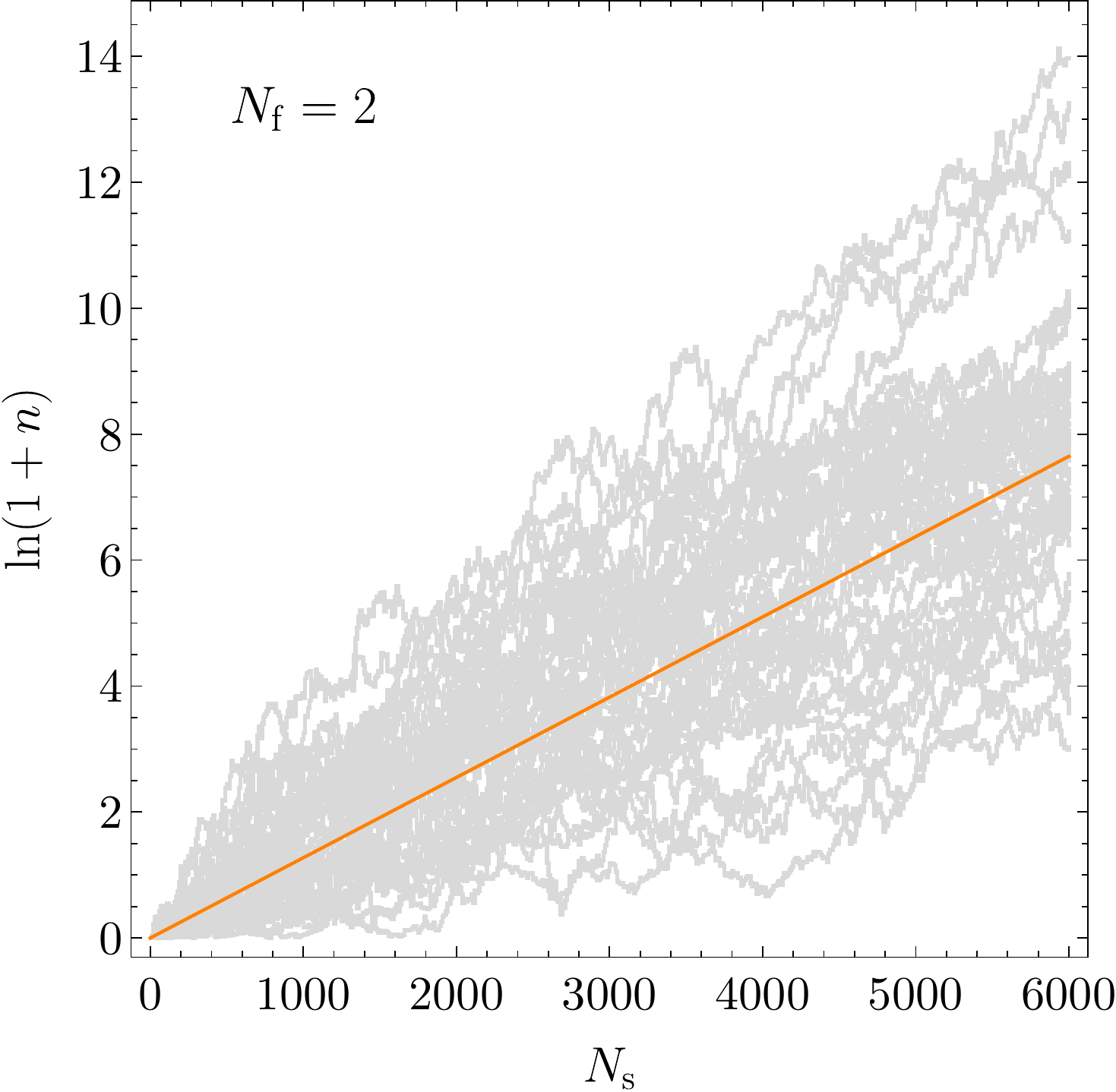}\label{fig:2trajectories1}}
    \hfill
    \subfloat{\includegraphics[width=0.49\textwidth]{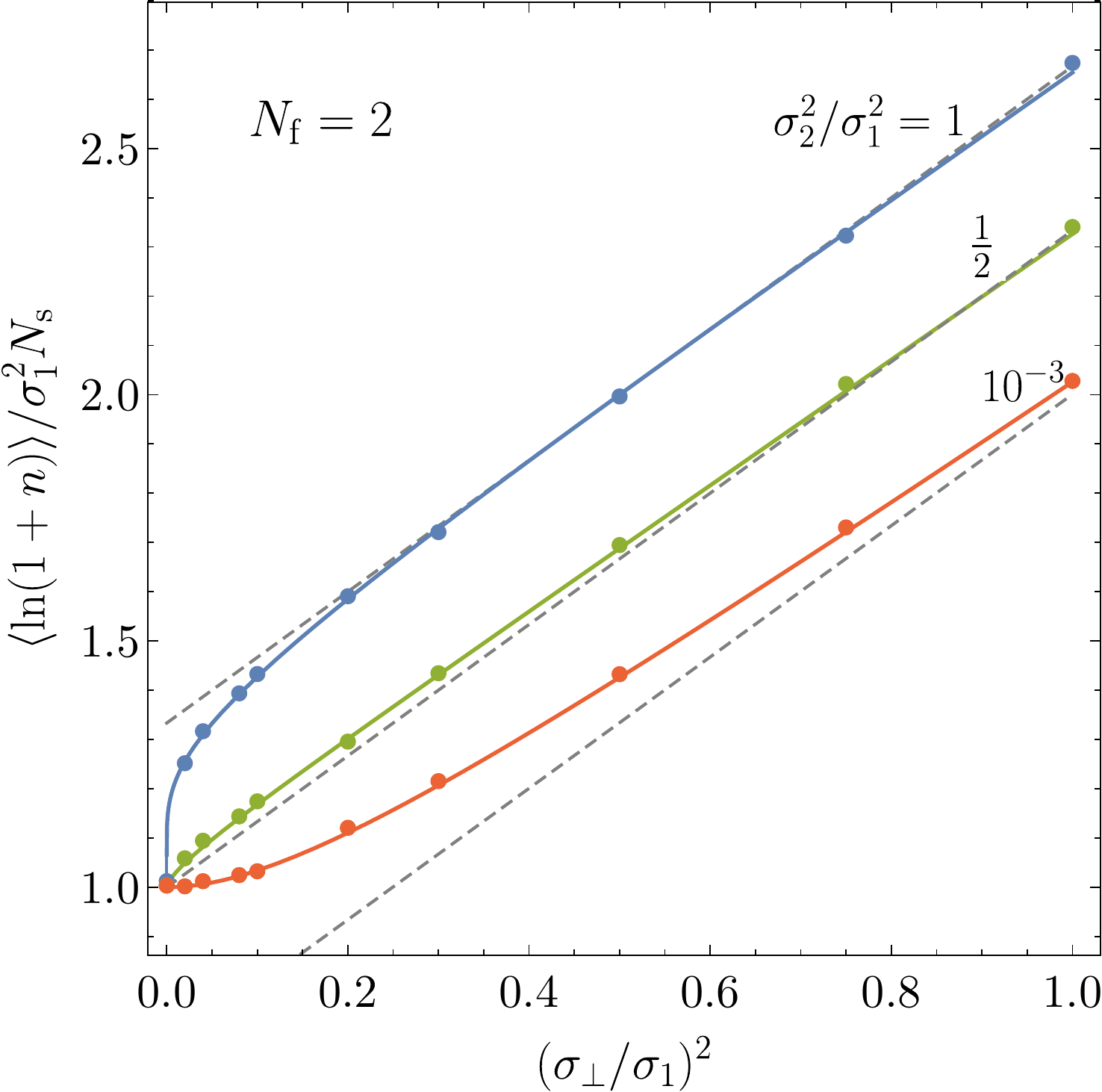}\label{fig:2trajectories2}}
    \caption{Left: Evolution of the occupation number per mode as a function of the number of scatterings $\Ns$ in the two-field case. Each gray line shows the numerical evolution for a particular realization of the scattering strengths and locations $\L_j$, $\tau_j$, with $\sigma_2^2=\sigma_1^2/10$ and $\sigma_{\perp}^2=\sigma_1^2/4$. The orange line is our analytic prediction for the typical occupation number. Right: Change in the typical occupation number for two fields as the number self and cross-couplings of the fields are varied. The solid lines are the theoretical prediction, the dots are the numerical results and the dashed lines correspond to the expectation from the {\it maximum entropy ansatz}-based trace formula.
        }
    \label{fig:2trajectories}
\end{figure}

Although we have now at hand all coefficients of the FP equation (\ref{eq:FPlambda}), we will not attempt its solution. Instead, we will focus on the calculation of the typical particle production rate at late times (see eq.~\eqref{eq:Fngen}), when it is assumed to have saturated to a constant value. 
In the limit of $\tau\rightarrow \infty$ (or equivalently $N_s\rightarrow \infty$) we expect that $n\gg 1$ and that a single eigenvalue
dominates the evolution at late times, $f_1\gg f_2$ or $f_2\gg f_1$. 
In the former case, using the coefficients we found in the previous section, we obtain
\Beq\label{eq:dtln2}
\lim_{\tau\rightarrow \infty}\partial_\tau \langle\ln(1+n)\rangle \rightarrow\left\langle l_1(\theta)-\frac{1}{2}\gamma_1(\theta)\right\rangle\,,
\Eeq
(the case $f_2\gg f_1$ is recovered by substituting $1\rightarrow 2$). The previous expression implies that, remarkably, the late-time particle production rate depends only on the probability distribution of the angular variable $\theta$. Therefore, to evaluate the right hand side, we need only the marginal distribution 
\beq
w(\theta) \;\equiv\; P(\theta,\tau\rightarrow\infty) \;=\; \int df_1\,df_2\, d\phi\, d\varphi\, d\psi\, P(\{f_1,f_2,\phi,\varphi,\psi\},\tau\rightarrow\infty)\,.
\eeq
%
To calculate $w(\theta)$ we need to solve the Fokker-Planck equation (\ref{eq:FPlambda}), which in the $\tau \rightarrow \infty$ limit can be approximated as stationary. Integrating both sides of eq.~(\ref{eq:FPlambda}) over $\{f_1,f_2,\phi,\varphi,\psi\}$ one finds the differential equation satisfied by the marginal probability density $w(\theta)$,
\beq\label{FPN}
\partial_{\theta}(c_1w) - \partial_{\theta}^2(c_{11}w)=0\,,
\eeq
where the coefficient functions are given by 
\begin{alignat}{2} \notag
c_1 &\equiv \lim_{\tau\rightarrow\infty} \langle \delta\theta^{(2)}\rangle  &&= -\frac{1}{4}\sin(2\theta)\left[ \sigma_1^2+\sigma_2^2 -2(3+\cot^2\theta)\sigma_{\perp}^2  \right]  - {\rm sgn}(\Delta f)(\sigma_1^2 - \sigma_2^2) \sin\theta\\ \label{eq:c1N21}
& &&\quad\, + 2\left[ \Theta(\Delta f)\gamma_4(\theta) - \Theta(-\Delta f)\gamma_5(\theta)\right]+ \frac{1}{4}\sin(2\theta)\csc^2\theta \gamma_6(\theta)\\ \label{eq:c1N22}
& &&= \frac{1}{4}\sin\theta(\cos\theta-2)\sigma_{\rm max}^2 + \frac{1}{4}\sin\theta(\cos\theta + 2)\sigma_{\rm min}^2 + \frac{1}{2}\cot\theta(3+\cos(2\theta))\sigma_{\perp}^2\,,
\end{alignat}
and
\beq
c_{11}  \equiv \frac{1}{2} \lim_{\tau\rightarrow\infty} \langle\delta\theta^{(1)}\delta\theta^{(1)}\rangle  = \frac{1}{4}(\sigma_1^2 + \sigma_2^2-4\sigma_{\perp}^2)\sin^2\theta + 2\sigma_{\perp}^2\,.
\eeq
In the step going from eq.~(\ref{eq:c1N21}) to eq.~(\ref{eq:c1N22}) we have assumed level repulsion ${\rm sgn}(\Delta f) = {\rm sgn}(\Delta\sigma^2)$. This assumption is justified at the perturbative level as long as $\sigma_{\perp}^2\leq\sigma^2_{\rm max}$ where $\sigma_{\rm max}^2 = \max\{\sigma_1^2,\sigma_2^2\}$. With the coefficient functions at hand, one can solve the differential equation (\ref{FPN}). Integration yields the equation
\beq\label{FPNred}
c_1 w - \partial_{\theta}(c_{11} w) = j\,,
\eeq
where $j$ denotes the current in the $\theta$-direction. Expansion around $\theta=0$ reveals that the marginal distribution would behave as $w(\theta) \sim \theta - (j/2\sigma_{\perp}^2)\,\theta\ln\theta$, meaning that for $j\neq0$ the probability distribution becomes negative in $(0,\pi)$. Thus, the normalizable, positive definite marginal probability $w(\theta)$ is found by straightforward integration of eq.~(\ref{FPNred}) with $j=0$: 
\begin{align}
w(\theta) &=  \exp\left(\int d\theta\, \frac{c_1(\theta)-c_{11}'(\theta)}{c_{11}(\theta)}\right)\\[5pt] \label{dist-w}
& = \frac{1}{\mathcal{N}} \times \begin{cases} 
\displaystyle \frac{\left( \sin{\theta} \right) }{\left(Q \sin^2{\theta}+ 1 \right)} \left[\frac{ \sqrt{1+Q} + \sqrt{Q} \cos{\theta}}{ \sqrt{1+Q} - \sqrt{Q} \cos{\theta}}  \right]^{\nu}, \quad  \sigma_1^2 + \sigma_2^2 \ge 4 \sigma_\perp^2, \\
\displaystyle \frac{\left( \sin{\theta} \right)}{\left(Q \sin^2{\theta}+ 1 \right) } \, \exp\left[ 2 \, \nu \arctan\left( \sqrt{\frac{|Q|}{1+Q}} \cos{\theta}\right) \right], \quad \sigma_1^2 + \sigma_2^2 < 4 \sigma_\perp^2,  \end{cases} 
\end{align}
where for general values of the exponents $\nu$ the normalization constant $\mathcal{N}$ must be determined numerically. In the above expression
\Beq
Q = \frac{\sigma_1^2 + \sigma_2^2 - 4\sigma_\perp^2 }{8 \sigma_\perp^2},\qquad\nu  = \frac{|\sigma_1^2 - \sigma_2^2|}{8 \sigma_\perp^2 \sqrt{{|Q|}{(1+Q)}}}\,.
\Eeq
The $w(\theta)$ calculated above can now be used to evaluate the right hand side of eq.~\eqref{eq:dtln2}. In the right panel of figure~\ref{fig:2trajectories}, we show the two-field theoretical prediction for $\langle \ln(1+n)\rangle$ for different values of the ratios $\sigma_a/\sigma_1$ (solid lines), and compare them to the numerical results (points). For a description of the numerical method see Appendix~\ref{ap:nm}. Since $w(\theta)$ only depends on the ratio of $\sigma_a/\sigma_1$, in the vertical axis the particle production rate has been normalized with respect to $\sigma_1^2$. 

%
%

The left panel of fig.~\ref{fig:2trajectories} shows a numerical solution for the evolution of the occupation number $n$ for an ensemble of fifty different realizations of randomly generated scatterers, verifying that $n$ executes a drifting random walk. The analytic result with $\sigma_2^2=\sigma_1^2/10$ and $\sigma_{\perp}^2=\sigma_1^2/4$ is shown as a orange line, and is clearly a good fit for the instantaneous average occupation number. Even more remarkable is the agreement between the numerical and analytic results in the right panel of fig.~\ref{fig:2trajectories}. For comparison, we have also provided the simple MEA inspired predictions, shown as the diagonal dashed gray curves. From eq.~(\ref{eq:meac}) with $\Nf=2$, the gray curves are given by
\beq
\partial_{\tau}\langle \ln(1+n)\rangle_{\rm MEA} = \frac{2}{3}\left(\sigma_1^2+\sigma_2^2 + 2\sigma_{\perp}^2\right)\,.
\eeq
It is clear from fig.~\ref{fig:2trajectories} that, for $\sigma_1^2=\sigma_2^2=2\sigma_{\perp}^2$ (equivalently all $\sigma^2_{ab}$ being the same), our calculation and the MEA based agree exactly. This is to be expected based on our proof in  section \ref{sect:MEA}. Indeed, in the $\L$-isotropic limit $\sigma_1^2=\sigma_2^2=2\sigma_{\perp}^2$ , the typical occupation number (\ref{eq:dtln2}) trivially reduces to the MEA result,
\beq
\lim_{\tau\rightarrow \infty} \partial_{\tau}\langle \ln(1+n)\rangle \;\xrightarrow{\sigma_1^2=\sigma_2^2=2\sigma_{\perp}^2}\; 2\sigma^2 \,.
\eeq
Moreover, in this all $\sigma^2_{ab}$ equal limit, the angular dependence of the late-time probability density is given simply by
\beq
dP(\phi,\theta,\varphi,\psi)  \;\xrightarrow{\sigma_1^2=\sigma_2^2=2\sigma_{\perp}^2}\; \frac{1}{2} \sin\theta\,d\phi\,d\theta\,d\varphi\,d\psi \;=\; d\mu({U(2)})\,,
\eeq
where $d\mu(U(2))$ denotes the invariant Haar measure for $U(2)$. This $U(2)$-`flat' probability density is also consistent with the MEA assumption of isotropy in the fundamental strip. Therefore, the MEA result is fully recovered assuming {\em only} $\L$-isotropy. 
%

It is worth discussing the agreement/difference between the MEA based results and the current calculation further. We note that the MEA based (top) dashed gray line is `tangential' to the actual analytical result, implying that its range of validity extends significantly beyond the values $2\sigma_\perp^2=\sigma_{1}^2=\sigma_2^2$. Also note that looking at the second from the top gray dashed line, the deviation of the actual result from the MEA based is small unless $\sigma_\perp$ is significantly different from $\sigma_1,\sigma_2$. These observations show that MEA-inspired ansatz (\ref{eq:meac}) is a good approximation to the exact result in the limit when all scattering strengths are comparable, but when they differ significantly care must be taken in using the approximation (at least in the two field case).

We also observe that, although there exist other special values of $\sigma_a^2$ (for example, large $\sigma^2_\perp$) for which the MEA based results intersect with the results of this section, they do not correspond to special points (i.e.~points of enhanced symmetry). We also note that the asymptotic behavior of the analytical curves in fig.~\ref{
fig:2trajectories}, which for $\sigma_{\perp}^2\gg \sigma_{1}^2,\sigma_2^2$ results in $\partial_{\tau}\langle \ln(1+n)\rangle \sim 4\sigma_{\perp}^2/\pi$, is different from the MEA prediction $\partial_{\tau}\langle \ln(1+n)\rangle \sim 4\sigma_{\perp}^2/3$. So the MEA lines should not necessarily  be interpreted as asymptotic curves in the large $\sigma_\perp$ limit.\footnote{It is unclear to us whether this regime is allowed by our approximations. Numerically we found that when $\sigma_\perp^2\gtrsim 10 \sigma_1^2$, we start seeing deviations from numerical simulations. This large $\sigma_\perp$ regime might violate our assumption that ${\rm sgn}(\Delta f) = {\rm sgn}(\Delta\sigma^2)$.}

\subsection{Summary}
Under the weak scattering approximation, we have solved the two field case in general without using the MEA, or $\L$-isotropy. As was to be expected from our proof in section \ref{sect:MEA}, we found precise agreement with MEA results when we have $\L$-isotropy and deviations from it otherwise. We verified that the MEA based trace formula for the typical occupation number (eq.~\eqref{eq:meac}) provides a good approximation to the more accurate calculation when the strengths of the self and cross couplings are all comparable to each other. Our analytic expressions (up to quadratures) here match the results from numerical simulations, and correctly capture the deviations from the MEA result.

\section{Beyond MEA: Generalization to Many Fields}
\label{sect:manyfields}

The analysis introduced in section~\ref{sect:twofield} for two fields can be extended to an arbitrary number of fields in a straightforward way, if the parametrization of the $\Nf\times \Nf$ unitary matrix $\u$ is known explicitly. However, for large $\Nf$ the computational expense increases dramatically, as now $(\Nf^2+\Nf)(\Nf^2+\Nf+3)/2$ correlators are necessary to solve the Fokker-Planck equation (\ref{eq:FPlambda}) for the probability density $P$. Therefore, in order to obtain a result that will allow us to draw generic conclusions for any number of fields, we will make the following simplifying assumptions: all off-diagonal couplings are equally distributed, and only one of the $\Nf$ fields has a statistically different self-coupling; that is
\begin{alignat}{2} \notag
&\sigma^2_{ab} \equiv \sigma_1^2/2\,,  \qquad &&a= b = \Nf \,,\\ \label{eq:NfieldAssu}
&\sigma^2_{ab} \equiv \sigma_2^2/2\,, &&  a = b \neq \Nf \,,\\ \notag
&\sigma^2_{ab} \equiv \sigma_{\perp}^2\,, && a\neq b\,.
\end{alignat}
Here we have associated $\sigma_1$ with the $\Nf$-th field for convenience; recall that the extra factors of $1/2$ for the diagonal entries stem from the definition in eq.~(\ref{eq:LambdaSigmaDef}). In what follows we will show that under the previous assumptions, and some seemingly reasonable conditions, only four correlators are necessary to calculate the typical occupation number in the general case. 

\subsection[Calculation of the Correlators]{Calculation of the Correlators \protect\footnote{In this section we denote the total number of fields by $N$ instead of $\Nf$ for notational simplicity.} }\label{subsec:Ncorr}

Let us explicitly calculate first the coefficients of the FP equation that depend on the first-order correction to the eigenvalues $f_i$. Luckily, the hardest part of the work has already been carried out in section~\ref{sect:MEA}, which led to the expression (\ref{varf1}) for the correlator $\langle \delta f_a^{(1)}\delta f_b^{(1)}\rangle$. Let us then consider the late time limit $N_s\rightarrow \infty$, in which, as in the two-field scenario, we expect one eigenvalue $f_a$ to be overwhelmingly dominant. We will identify this dominant eigenvalue with the $N$-th one, $f_N\gg \{f_a,1\}$. In this limit the sum $\sum_{a,b=1}^{N}\langle \delta f_a^{(1)}\delta f_b^{(1)}\rangle_{\delta \tau}/\delta\tau$ will be dominated by its last term, which in turn, under the assumption (\ref{eq:NfieldAssu}), can be approximated as 
\beq\label{eq:dfdf}
\langle \delta f_N^{(1)}\delta f_N^{(1)}\rangle  \stackrel{\infty}{=} 2f_{N}^2 \left[  \sigma_1^2\, |u_{N N}|^4  + \sigma_2^2 \sum_{a=1}^{N-1}|u_{aN}|^4 + 2\sigma_{\perp}^2 \sum_{\substack{a,b=1\\  a\neq b}}^{N}|u_{aN}|^2|u_{bN}|^2\right]\,.
\eeq
In the previous expression $\stackrel{\infty}{=}$ denotes an equality valid in the $N_s\rightarrow \infty$ regime.

 To go further, an explicit parametrization of a generic $N\times N$ unitary matrix is needed. In analogy with the two field case, it is convenient to express the unitary matrix $\u$ in terms of the (generalized) Euler angular variables. As outlined in \cite{TilmaSudarshanSUN}, an Euler parametrization of an element of $SU(N)$ can be iteratively constructed as follows:\footnote{It is sufficient to restrict ourselves to the $SU(N)$ subgroup of $U(N)$ as the extra phase is canceled in the products appearing on (\ref{eq:dfdf}).} 
\beq
{\bsf u} = \left(\prod_{2 \leqslant k \leqslant N} {\bsf A}(k)\right)\cdot [SU(N - 1)]\cdot e^{i {\boldsymbol \lambda}_{N^2-1}\alpha_{N^2-1}}\,,\qquad {\bsf A}(k) = e^{i {\boldsymbol \lambda}_3 \alpha_{(2k-3)}}e^{i {\boldsymbol \lambda}_{(k-1)^2+1}\alpha_{2(k-1)}}\,,
\eeq
where the relevant generators of the $su(N)$ Lie algebra are given by
\begin{alignat}{2}  \displaybreak[0]
{\boldsymbol \lambda}_{3} &= \frac{1}{2}\left(
\begin{matrix}
1 & 0 & \cdots & 0\\
0 & -1 & \cdots & 0\\
\vdots & \vdots & \ddots& \vdots\\
0 & 0 & \cdots & 0
\end{matrix}
\right)_{N\times N}\,, \qquad && 
{\boldsymbol \lambda}_{(k-1)^2+1} = \frac{1}{2}\left(
\begin{matrix}
\left[
\begin{matrix}
0 & \cdots & -i\\
\vdots & \ddots & \vdots\\
i & \cdots & 0
\end{matrix}
\right]_{k\times k} & \cdots & 0\\
\vdots &  \ddots& \vdots\\
0&\cdots & 0
\end{matrix}
\right)_{N\times N}\ (k<N)\,, \\[5pt] \displaybreak[0]
{\boldsymbol \lambda}_{(N-1)^2+1} &= \frac{1}{2}\left(
\begin{matrix}
0 & 0 & \cdots & -i\\
0 & 0 & \cdots & 0\\
\vdots & \vdots & \ddots& \vdots\\
i & 0 & \cdots & 0
\end{matrix}
\right)_{N\times N}\,, &&
{\boldsymbol \lambda}_{N^2-1} = \sqrt{\frac{1}{2N(N-1)}} \left(
\begin{matrix}
1 & 0 & \cdots & 0\\
0 & 1 & \cdots & 0\\
\vdots & \vdots & \ddots& \vdots\\
0 & 0 & \cdots & -(N-1)
\end{matrix}
\right)_{N\times N}.
\end{alignat}
This parametrization implies that the magnitudes of the elements of the last column of $\u$ can be written as
\beq\label{uiN}
|u_{aN}| = \left(
\begin{matrix}
\cos({\alpha_2}/2)\cos(\alpha_4/2)\cdots\cos(\alpha_{2(N-3)}/2) \cos(\alpha_{2(N-2)}/2) \sin(\theta/2)\\[5pt]
\cos({\alpha_4}/2)\cos(\alpha_6/2)\cdots\cos(\alpha_{2(N-2)}/2) \sin(\alpha_{2}/2) \sin(\theta/2)\\[5pt]
\cos({\alpha_6}/2)\cos(\alpha_8/2)\cdots\cos(\alpha_{2(N-2)}/2) \sin(\alpha_{4}/2) \sin(\theta/2)\\[5pt]
\vdots\\[5pt]
\cos(\alpha_{2(N-2)}/2) \sin(\alpha_{2(N-3)}/2) \sin(\theta/2)\\[5pt]
 \sin(\alpha_{2(N-2)}/2) \sin(\theta/2)\\[5pt]
 \cos(\theta/2)
\end{matrix}
\right)\,,
\eeq
where, in order to facilitate the comparison with the two-field result, we have denoted $\alpha_{2(N-1)}\equiv \theta$. Therefore, the dominant correlator in eq.~(\ref{eq:dfdf}) evaluates to
\begin{align}\notag 
\langle \delta f_N^{(1)}\delta f_N^{(1)}\rangle \stackrel{\infty}{=}  2f_{N}^2 \Big[  &\sigma_1^2 \cos^4(\theta/2)  + \sigma_2^2 \sin^4(\theta/2)\, \Fo (\Omega_{N})\\ \label{eq:dfdfNf}
& + 2\sigma_{\perp}^2 \big(1-\cos^4(\theta/2)-\sin^4(\theta/2) \Fo(\Omega_{N})\big) \Big]\,.
\end{align}
Here $\Fo(\Omega_{N})$ is a function of the Euler angles other than $\theta$ appearing in eq.~(\ref{uiN}). It can be calculated iteratively as
\beq\label{eq:Fodef}
\Fo(\Omega_N) = \Fo(\Omega_{N-1})\cos^4(\alpha_{2(N-2)}/2) + \sin^4(\alpha_{2(N-2)}/2)\,.
\eeq
with $\Fo(\Omega_2)=1$. In what follows we will denote $\Fo(\Omega_N)$ by $\Fo_{\Omega}$. \\

The general expression for the second-order change $\delta f_a^{(2)}$ is given by eq.~(\ref{varf2}). Specializing to the particular case with eq.~(\ref{eq:NfieldAssu}), in the late time limit it is straightforward to show that the perturbations to the non-dominant eigenvalues are negligible, $\langle \delta f_a^{(2)}\rangle \stackrel{\infty}{=} 0$ for $a\neq N$, while for the dominant eigenvalue we have
\begin{align}\notag
\langle \delta f_N^{(2)}\rangle &\;\stackrel{\infty}{=}\;  2f_{N}\left[ \sigma_1^2\, |u_{N N}|^2  + \sigma_2^2 \sum_{a=1}^{N-1}|u_{aN}|^2 + 2\sigma_{\perp}^2 \sum_{\substack{a,b=1\\  a\neq b}}^{N} |u_{bN}|^2 \right]\\ \label{eq:d2fNf}
& \;=\; 2f_{N}\big[\sigma_1^2\cos^2(\theta/2) + \sigma_2^2\sin^2(\theta/2) + \sigma_{\perp}^2(N-1)\big]\,.
\end{align}\\

The first- and second-order corrections for the entries of $\u$ have been written in eqns.~(\ref{deltau1}) and (\ref{deltau2}). In principle then, the angular correlators can be evaluated by solving these two equations in terms of the angular perturbations; namely
\beq\label{eq:duda}
\delta{\bsf u} = \frac{\partial {\bsf u}}{\partial \alpha_i}\,\delta\alpha_i\,.
\eeq
In practice, the solution for this system of equations and the subsequent evaluation of the first and second order correlators cannot be done for a generic $N$ and must be performed on a case by case basis (c.f. the two-field calculation). Nevertheless, for the $\theta$ correlators,  eqns.~(\ref{deltau1}), (\ref{deltau2}) and (\ref{uiN}) can be used to relate $\delta u_{N N}$ and $\delta\theta$. In complete analogy with the previous calculation for the $\delta f$ expectation values, in the $f_{N}\gg f_a$ limit one obtains 
\begin{align} \notag
\langle\delta \theta^{(1)}\delta\theta^{(1)}\rangle &= 
4\csc^{2}\theta \left( u_{NN}^{*2} \langle \delta u^{(1)}_{NN} \delta u^{(1)}_{NN}\rangle + 2|u_{NN}|^2 \langle \delta u^{(1)}_{NN} \delta u^{*(1)}_{NN}\rangle + u_{NN}^2 \langle \delta u^{*(1)}_{NN} \delta u^{*(1)}_{NN}\rangle \right)\\ \label{eq:dthetadtheta}
&\stackrel{\infty}{=} \frac{1}{2}\left[\sigma_1^2 + \Fo_{\Omega} \sigma_2^2 - 2(1+\Fo_{\Omega})\sigma_{\perp}^2 \right]\sin^2\theta + 4\sigma_{\perp}^2\,,
\end{align}
and
\begin{align} \notag
\langle\delta \theta^{(2)}\rangle &=  -  \frac{1}{2}\cot\theta\, \langle\delta \theta^{(1)}\delta\theta^{(1)}\rangle - 2\csc\theta \left( u_{NN}^{*} \langle \delta u^{(2)}_{NN} \rangle + u_{NN} \langle \delta u^{*(2)}_{NN} \rangle + \langle \delta u_{NN}^{*(1)}\, \delta u_{NN}^{*(1)} \rangle \right) \\[5pt] \notag
&\stackrel{\infty}{=}\frac{1}{4}\, \sigma_1^2 \sin\theta(\cos\theta-2) +\frac{1}{4}\,\sigma_2^2 \sin\theta\big(\Fo_{\Omega}\cos\theta + 2(2-\Fo_{\Omega}) \big)  \\ \label{eq:dtheta2}
&\quad\ + \frac{1}{2}\sigma_{\perp}^2\Big[ 4(N-2)\csc\theta + 4(N-1)\cot\theta -\sin\theta\big((1+\Fo_{\Omega})\cos\theta + 2(1-\Fo_{\Omega})\big)  \Big]\,.
\end{align}

\subsection{The Typical Occupation Number}

In the previous section we have shown that under the assumptions stated in eqns. (\ref{eq:NfieldAssu}) we can explicitly calculate the eigenvalue correlators in the $\tau \rightarrow \infty$ limit, as well as the set of correlators for one angular variable. Following the same argument for the two-field case discussed in section~(\ref{sec:typical2}), we can calculate the typical occupation number in the $\Nf$-field case by considering the late-time limit of (\ref{eq:Fngen}). Substitution in eq.~(\ref{eq:Fngen}) of the expressions (\ref{eq:dfdfNf}) and (\ref{eq:d2fNf}) we obtain
\beq\label{eq:typicalN}
\partial_{\tau}\langle \ln(1+n)\rangle \;\stackrel{\infty}{=}\; \left\langle l (\theta) - \frac{1}{2} \gamma (\theta,\Omega_{\Nf})  \right\rangle\,,
\eeq
where the functions $l(\theta)$ and $\gamma(\theta,\Omega_{N})$ are the generalization of their two-field counterparts,
\begin{alignat}{2}
l(\theta) &= 2 \big[\sigma_1^2\cos^2(\theta/2) &&+ \sigma_2^2\sin^2(\theta/2) + \sigma_{\perp}^2(\Nf-1)\big]\,,\\ \notag
\gamma(\theta,\Omega_{\Nf}) &= 2 \Big[  \sigma_1^2 \cos^4(\theta/2)  &&+ \sigma_2^2 \sin^4(\theta/2)\, \Fo_{\Omega}\\
& &&+ 2\sigma_{\perp}^2 \big(1-\cos^4(\theta/2)-\sin^4(\theta/2)\, \Fo_{\Omega}\big) \Big]\,.
\end{alignat}
Note first that, as expected, in the isotropic-$\L$ limit, the left-hand side of eq.~(\ref{eq:typicalN}) is angle-independent, and  the typical occupation number takes the MEA value consistent with eq.~\eqref{eq:meac}:
\beq
\partial_{\tau}\langle \ln(1+n)\rangle \;\xrightarrow{\sigma_1^2=\sigma_2^2=2\sigma_{\perp}^2}\; \sigma^2\Nf\,.
\eeq
As a second particular example, in the large $\Nf$-limit, the angular dependence of the expression inside the angular brackets in eq.~(\ref{eq:typicalN}) becomes subdominant. Therefore, one obtains the universal limit
\beq\label{eq:largenf}
\partial_{\tau}\langle \ln(1+n)\rangle \;\xrightarrow{\Nf,\Ns\gg 1}\; 2\sigma_{\perp}^2\Nf\,.
\eeq
In the general case, evaluation of eq.~(\ref{eq:typicalN}) requires the knowledge of the late-time marginal probability density which is a function of {\em all} the angles that parametrize (\ref{uiN}), due to the explicit angular dependence of $\Fo_{\Omega}$. Analogously to the two-field case, such function could in principle be found for a given $\Nf$ after the inversion of eq.~(\ref{eq:duda}) and the computation of all angular correlators. We have not attempted to take on such a task, and instead we will go beyond MEA by calculating the right-hand side of eq.~(\ref{eq:typicalN}) in complete analogy with the two-field case. Namely, we will {\em assume} that in the stationary $N_s\rightarrow \infty$ limit, the dependence on angles other than $\theta$ of the probability density is factorizable. 
The compact nature of the angular variables and the previous assumption imply that, similarly to the two-field case, the required marginal distribution $w(\theta)$ may be obtained from the solution of the single-variable Fokker-Planck equation (\ref{FPN}), where the coefficient functions $c_1$ and $c_{11}$ are now determined from the expectation values noted in eqns.~(\ref{eq:dthetadtheta}) and (\ref{eq:dtheta2}), and correspond to
\begin{align}\notag
c_1 &= \frac{1}{4} \cos\theta \sin\theta \,\big[ \sigma_1^2 +  \Fo \sigma_2^2  - 2(1+\Fo)\sigma_{\perp}^2 \big]   - \frac{1}{2}\sin\theta\, \big[ \sigma_1^2 + (\Fo-2)\sigma_2^2 + 2(1-\Fo)\sigma_{\perp}^2\big]\\ \displaybreak[0]
&\qquad + 2\csc\theta\,(\Nf-2)\sigma_{\perp}^2  + 2 \cot\theta\,(\Nf-1)\sigma_{\perp}^2 \,,\\[5pt]  
c_{11}&= \frac{1}{4}\left[\sigma_1^2 + \Fo \sigma_2^2 - 2(1+\Fo)\sigma_{\perp}^2 \right]\sin^2\theta + 2\sigma_{\perp}^2  \,.
\end{align}
In the previous expressions, we have introduced the notation $\langle \Fo_{\Omega}\rangle=\Fo$. To be able to go further, we need the explicit form of this expectation value. Although in general this can only be achieved by solving the full FP equation and averaging, remarkably, in the $\u$-isotropic limit, a closed-form for the expectation value of the angular function $\Fo_{\Omega}$ can be calculated; for the interested reader the details are shown in appendix~\ref{ap:Fav}. It results in the following,
\beq\label{eq:fodef}
\langle \Fo_{\Omega}\rangle_{\rm Haar} = 2/\Nf\,.
\eeq
We will then consider the additional simplifying approximation $\Fo=\langle \Fo_{\Omega}\rangle_{\rm Haar}$. Despite the fact that such approximation is in general unjustified, we expect it to lead to reasonable results in the regime $\sigma_1^2\sim \sigma_2^2\sim \sigma_{\perp}^2$. With these assumptions, the vanishing-current solution of eq.~(\ref{FPN}) is given by
\begin{align}\label{eq:analnf}
w(\theta) &=  \exp\left(\int d\theta\, \frac{c_1(\theta)-c_{11}'(\theta)}{c_{11}(\theta)}\right)\\ \label{eq:Nfieldsol1}
& =  \frac{1}{\mathcal{N}}\,\big(\tan(\theta/2)\big)^{\Nf-2}\,\frac{(\sin\theta)^{\Nf-1}}{(Q\sin^2\theta+1)^{\Nf/2}}\,
\exp\left[2 \nu\, {\rm arctanh}\left(\sqrt{\dfrac{Q}{1+Q}}\,\cos\theta\right)\right]\, ,
\end{align}
where now 
\begin{align}
Q &\;=\;\frac{\sigma_1^2+\Fo\sigma_2^2 - 2(1+\Fo)\sigma_{\perp}^2}{8\sigma_{\perp}^2}\,,\\[5pt]
\nu &\;=\; \frac{\Nf\, \sigma_1^2 + (\Fo\Nf-4) \sigma_2^2 + 2(4-\Nf-\Fo\Nf)\sigma_{\perp}^2 }{16\sigma_{\perp}^2\sqrt{Q(1+Q)}}\,.
\end{align}
Here $\mathcal{N}$ denotes the normalization constant, which must be calculated numerically in general. For negative values of $Q$ the expression (\ref{eq:Nfieldsol1}) is to be understood in the analytically continued sense. Note that in the $\Nf\rightarrow2$ limit, the solution reduces to the one in eq.~(\ref{dist-w}), as expected. The $w(\theta)$ calculated above can now be used in the right hand side of eq.~\eqref{eq:typicalN} to calculate the typical occupation number.
%
%
\begin{figure}[ht]
\centering
    \subfloat{\includegraphics[width=0.49\textwidth]{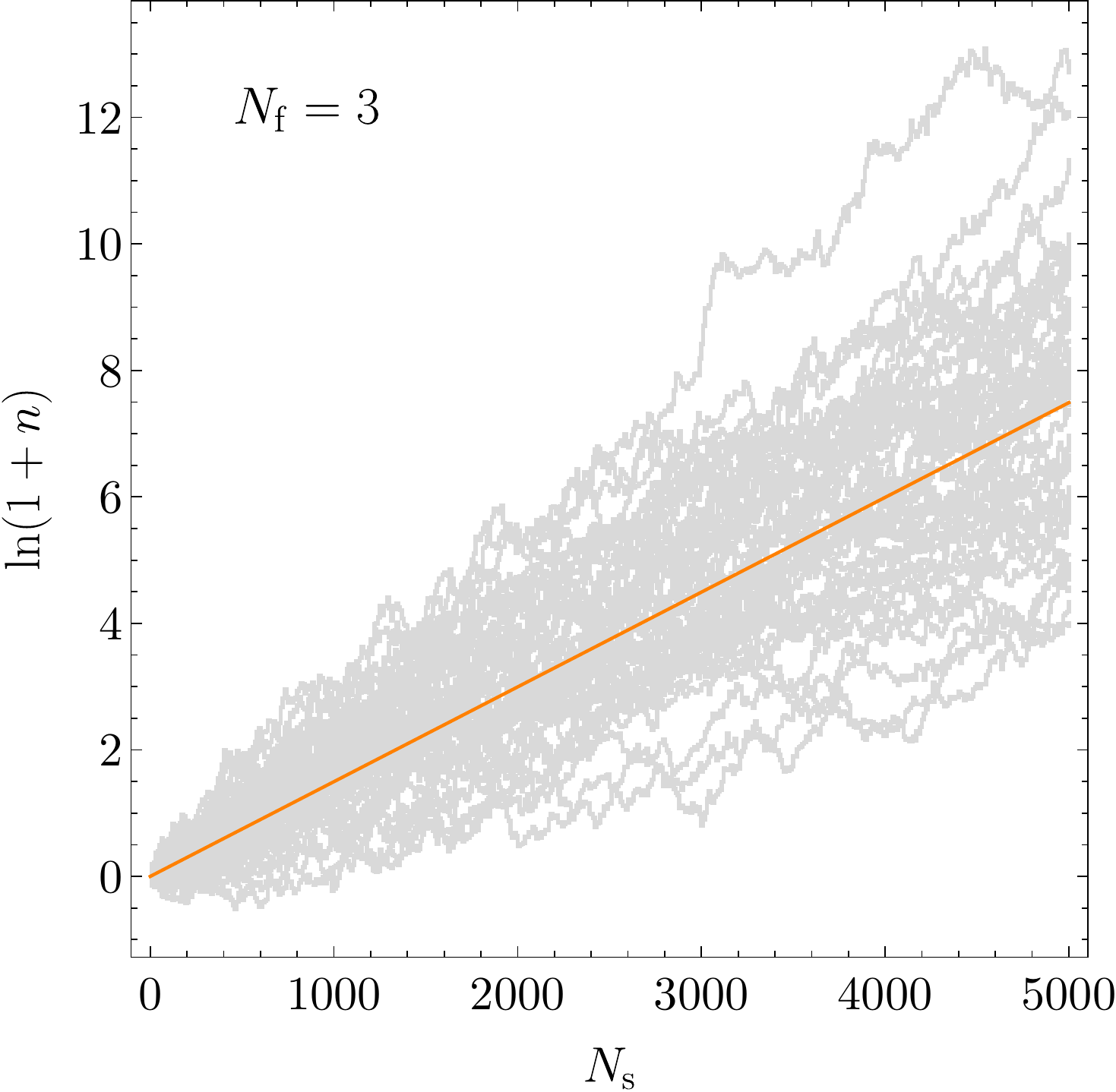}\label{fig:3trajectories1}}
    \hfill
    \subfloat{\includegraphics[width=0.49\textwidth]{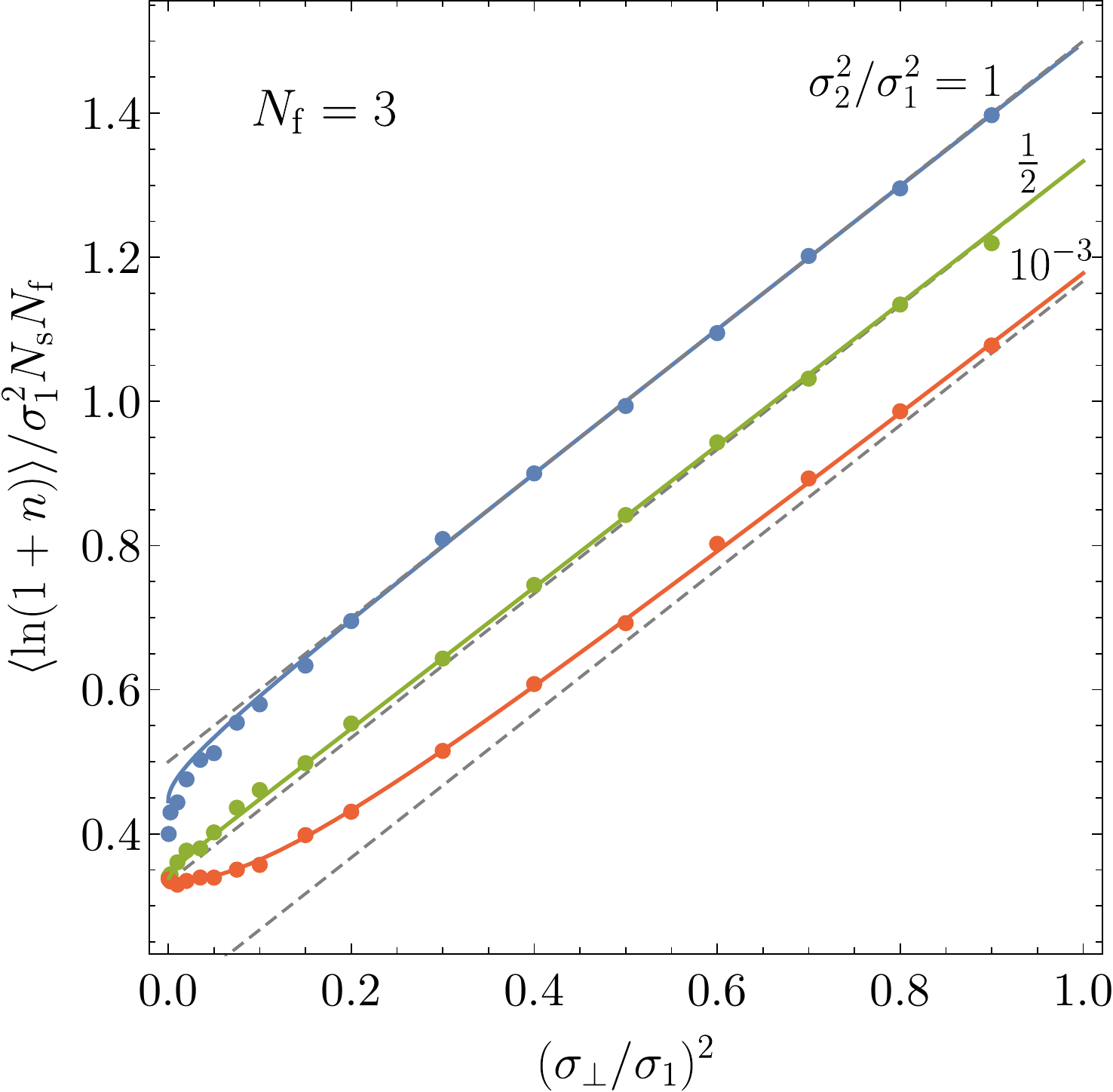}\label{fig:3trajectories2}}\\
    \subfloat{\includegraphics[width=0.49\textwidth]{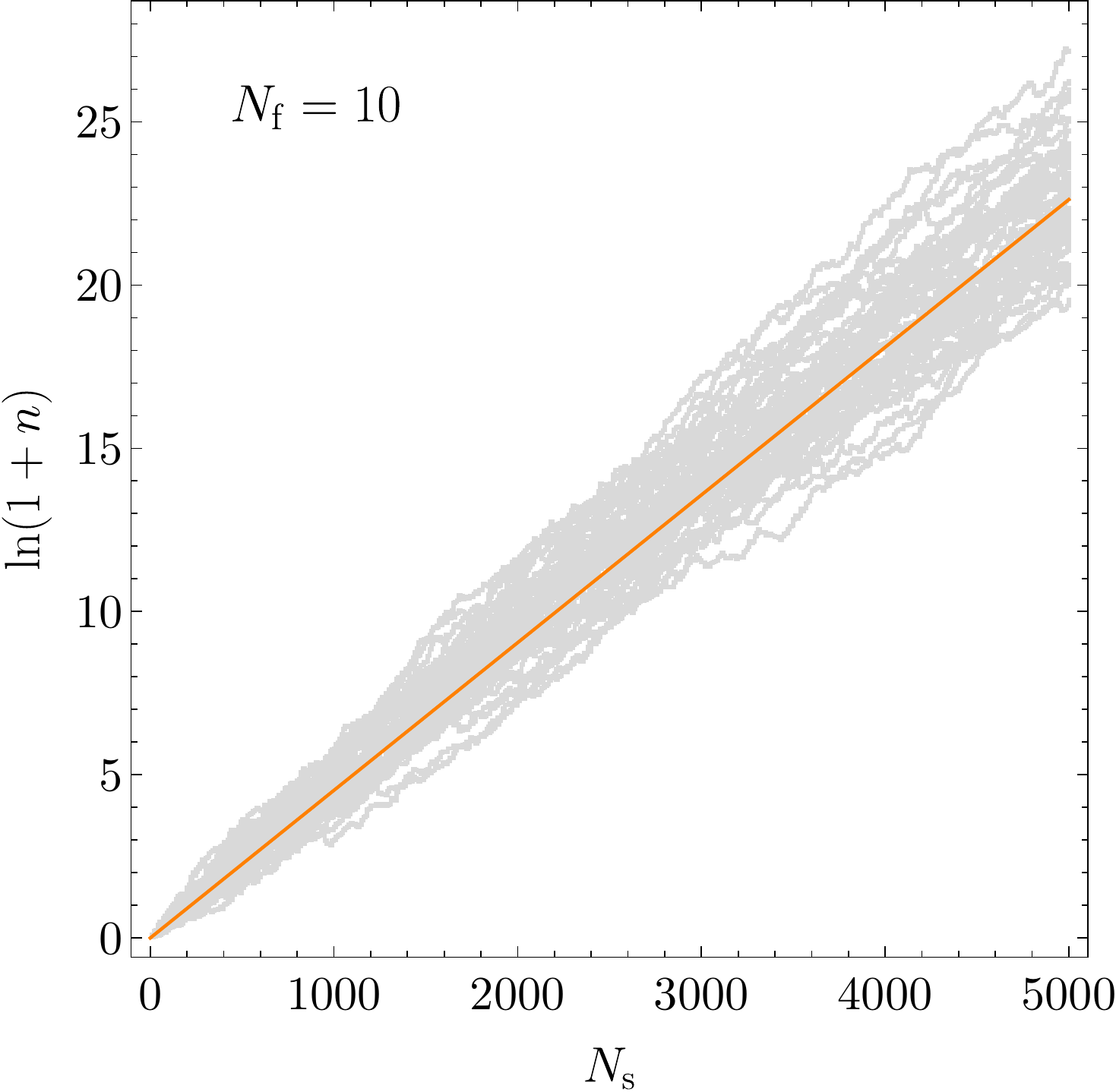}\label{fig:10trajectories1}}
    \hfill
    \subfloat{\includegraphics[width=0.49\textwidth]{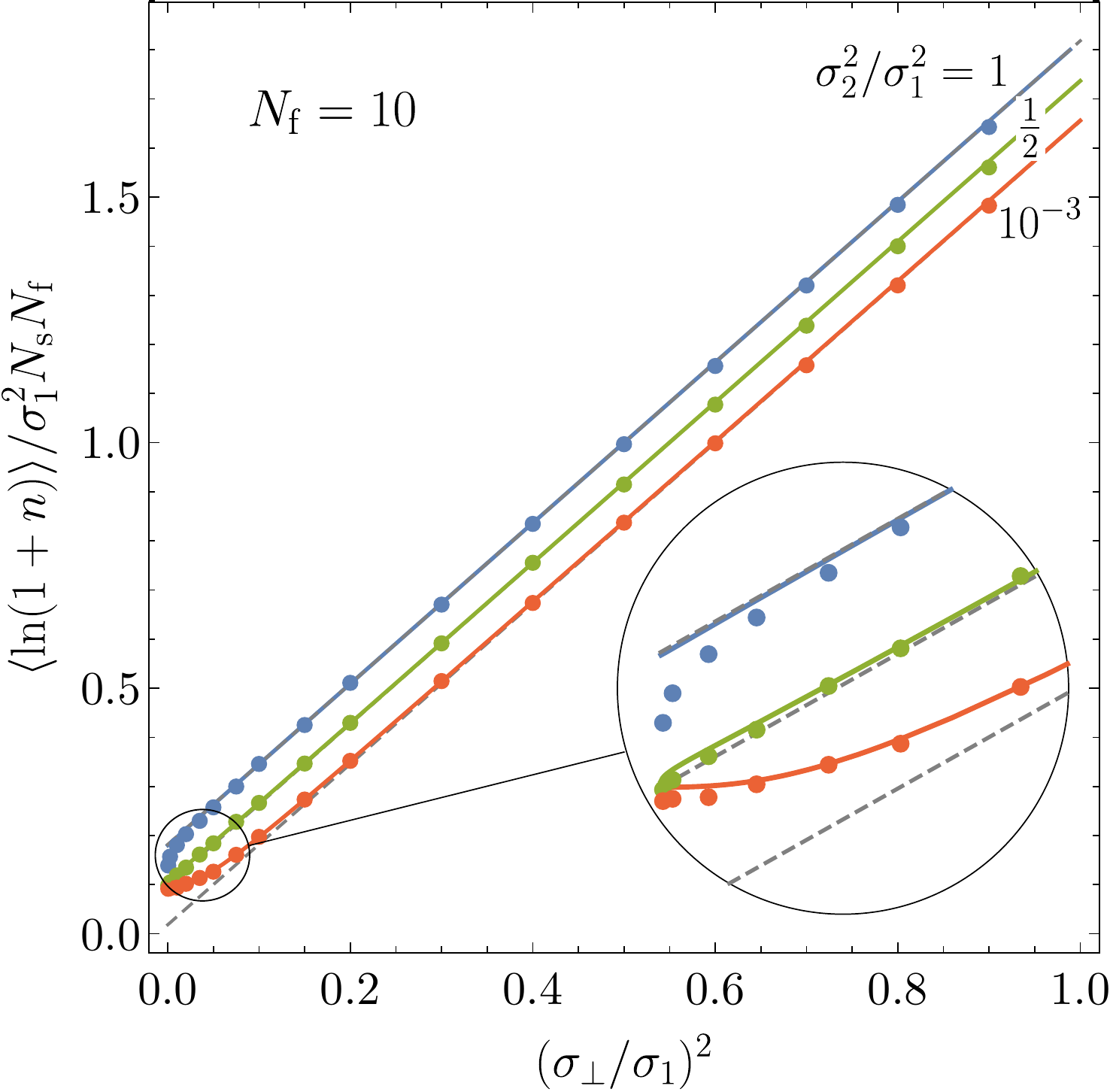}\label{fig:10trajectories2}}
    \caption{Left: Evolution of the occupation number per mode as a function of the number of scatterings $\Ns$ for $\Nf=3$ (top) and $\Nf=10$ (bottom). Each gray line shows the numerical evolution for a particular realization of the scattering strengths and locations $\L_j$, $\tau_j$, with $\sigma_2^2=\sigma_1^2/10$ and $\sigma_{\perp}^2=\sigma_1^2/4$. The orange line is our analytic prediction for the typical occupation number. Right: Typical occupation number as a function of the self and cross-coupling strengths for $\Nf=3$ (top) and $\Nf=10$ (bottom) interacting fields. The thick dots correspond to the numerical results, the continuous curves are the theoretical prediction with $\Fo=\langle\Fo_{\Omega}\rangle_{\rm Haar}$. The dashed gray curves are the MEA expectation. The region where our theoretical predictions deviate from the numerical ones (and from MEA-based results) shrinks as $1/\Nf$ for large $\Nf$. 
        }
    \label{fig:10trajectories}
\end{figure}

The numerical solution for the evolution of the occupation number $n$ for fifty different rea\-li\-za\-tions of randomly generated scatterers and $\Nf=3,10$ is shown in the left panels of figure~\ref{fig:10trajectories}. Similarly to the two-field case, the analytic result (shown as the orange line) accurately approximates the typical occupation number.
%

In the right panels of figure~\ref{fig:10trajectories} we show the three- and ten-field theoretical predictions for different values of the ratios $\sigma_a/\sigma_1$, and compare them to the numerical results. It must be noted that the vertical axis has been rescaled by $\Nf$ compared to the two-field case for a better reading of the results. In both cases, but more evidently in the ten-field case, the deviation between the theoretical prediction and the numerical results is significant in the regime $\sigma_{\perp}^2\ll \sigma_1^2$, $\sigma_2^2\sim \sigma_1^2$. We believe that this deviation is related to the breakdown of our assumptions regarding $\Fo$ and $w(\theta)$; namely, the factorizability of the angular dependence of the probability density and/or the naive assumption of a scattering strength-independence of $\Fo$. 
%
%
%

 The MEA ansatz by itself provides a good fit to the data in the $\sigma_1\sim\sigma_2\sim\sigma_{\perp}$ regime. Here the MEA solution can be written as
\beq
\label{eq:MEANfields}
 \partial_{\tau}\langle \ln(1+n)\rangle_{\rm MEA} \;\stackrel{\infty}{=}\;  \frac{2 }{\Nf+1}\left[\sigma_1^2 + (\Nf-1)\sigma_2^2 + \Nf(\Nf-1)\sigma_{\perp}^2\right]\,. 
\eeq
The corresponding curves are shown in fig.~\ref{
fig:10trajectories} (right) as the dashed gray lines.\\ 

Unlike the two-field case, for $\Nf>2$ fields we lack the knowledge of the full set of coefficients of the FP equation, and thus we cannot claim that, in general, $\L$-isotropy implies $\u$-isotropy, although we can prove that the distribution of the angle relevant for the typical occupation number takes the $\u$-isotropic form. 
In the $\L$-isotropic limit, the analytical solution in eq.~(\ref{eq:analnf}) evaluates to
\beq\label{eq:wthLiso}
w(\theta)\big|_{\text{$\L$-iso}} \propto \tan(\theta/2)^{\Nf-2}(\sin\theta)^{\Nf-1} \propto \cos(\theta/2)\sin(\theta/2)^{2\Nf-3}\,,
\eeq
which is nothing but the $U(\Nf)$-flat result $w(\theta)\big|_{\rm Haar}$, see Appendix~\ref{ap:Fav}. 

At any rate, for $\Nf\gg1$, based on eqns.~\eqref{eq:largenf} and \eqref{eq:MEANfields} we expect the full analytical result and the result based on the MEA to be coincident in most of the parameter space for a large number of fields, save for $\sigma_{\perp}^2\ll \sigma_1^2/\Nf$ 
 with $\sim \Nf^2$ contributions from $\sigma_\perp^2$ dominating the answer. In terms of fig.~\ref{
 fig:10trajectories}, the separation between the curves, and the magnified region are observed to shrink as $\sim 1/\Nf$.

For generic scattering strengths, not necessarily of the form in eq.~(\ref{eq:NfieldAssu}), it has been argued that, due to the concentration of measure property of the unitary group, the parameter space region for which the dependence of $\delta f$ deviates significantly from that predicted by MEA, becomes negligibly small as $\Nf \rightarrow \infty$~\cite{BachmannRoeck}. We therefore expect that the validity of the trace formula (\ref{eq:dthaar}) carries over beyond the assumption (\ref{eq:NfieldAssu}), as long as the disorder strengths do not greatly deviate from $\L$-isotropy.

\subsection{Summary}
We have derived the typical occupation number in the $\Nf>2$ case without appealing to the maximum entropy ansatz, but with some restrictions (see eq.~\eqref{eq:NfieldAssu}) on the form of the $\sigma_{ab}^2$. We had to make a couple of ad-hoc assumptions along the way: (1) $P(\theta,\alpha_2\hdots)=w(\theta)p(\alpha_2,\hdots)$. (2) $\Fo=\langle\Fo_{\Omega}\rangle_{\rm Haar}$. These were motivated by their validity within the MEA based approach. We found that the results then match the numerical simulations exceptionally well, apart from the region $\sigma_\perp^2\ll \sigma_1^2/\Nf$. The MEA based result (see eq.~\eqref{eq:MEANfields}) and the result from this section match quite well apart from a small region $\sigma_\perp^2\ll \sigma_1^2/\Nf$ which scales as $1/\Nf$. Thus in the large $\Nf$ limit, it is safe to use the simple MEA based formula (see eq.~\eqref{eq:meac}) for the typical occupation number.
\section{Summary and Future Outlook}
\label{sect:summary}
Non-adiabatic particle production during and after inflation can impact the generation of curvature perturbations as well the efficiency of energy transfer during reheating. In this work, we derived a general Fokker-Planck equation for the probability distribution of the total occupation number of $\Nf$ coupled fields under the assumption that the typical particle production rate per interaction is small. Our analytic calculation of the typical total occupation number based on the FP equation agreed exceptionally well with direct numerical computations in the limit of large number of interactions.

For analytic tractability, we restricted our attention to cases where the effective stochastic masses and couplings of the fields can be described by Dirac-delta function with amplitudes and locations drawn from different distributions.  Furthermore, we ignored effects of expansion and backreaction on the perturbations. Physically this corresponds to non-adiabatic interactions whose temporal width is small compared to the wavelength of the mode-functions of interest, and that we are restricting our attention to sub-horizon, linearized perturbations. 

In the present manuscript, we have extended the framework of \cite{Amin:2015ftc} for calculating the typical occupation number of $\Nf$ coupled fields. The key difference from \cite{Amin:2015ftc} was that we did not rely on the maximum entropy ansatz (MEA). We elucidated the domain of validity of the MEA in terms of the relative strengths of the effective masses and cross couplings of the fields ($\sigma_{ab}^2$) and the number of fields ($\Nf$). We proved that MEA based results coincide with our present results when all $\sigma^2_{ab}$ are equal. 
We verify that a simple and useful {\it trace formula} based on MEA agrees well with detailed numerical results when the $\sigma^2_{ab}$ are comparable (not necessarily equal), or more generally if $\Nf\gg 1$ even when some of the $\sigma^2_{ab}$ differ significantly. The deviations from this trace formula are captured by our more detailed framework here, albeit at an increased level of complexity in the derivation. 

In detail, we note that the expressions for the typical occupation number in the $\Nf=1$ and $\Nf=2$ are given in closed form (up to quadratures). To do the same for $\Nf>2$ fields, additional simplifying assumptions are needed. These assumptions, however, mattered less when the number of fields is large ($\Nf\gg 1$) where we find the MEA based {\it trace formula} to be a good approximation for a very wide range of effective masses and couplings.

The overall approach advocated here is to avoid relying on detailed model building, and rather take a coarse-grained approach to the particle production in the early universe. We believe that this is reasonable way forward given (i) the complexity of fundamental physics models of the early universe, and (ii) the simplicity of the existing data from the early universe. While the technical approach is different here, the EFT of inflation and reheating \cite{Cheung:2007st, Weinberg:2008hq,Ozsoy:2015rna,Giblin:2017qjp}, as well as disorder during inflation \cite{Green:2014xqa}, is based on a similar underlying philosophy. There is also similarity in the underlying view taken by the authors of \cite{McAllister:2012am, Dias:2012nf, Dias:2016slx}, where simple power spectra of the curvature perturbations are shown to exist in spite of the complexity of the underlying models.  

A possible means of moving away from the weak scattering limit is to use Random Matrix Theory (RMT), taking advantage of the large number of fields $\Nf$ and the large number of scatterings $\Ns$ \cite{crisanti1993products,oseledec1968multiplicative, 2005cond.mat.11622B}. 
In the case of rapid particle production, a detailed investigation of back-reaction and nonlinear mode effects might be needed after a few interactions. Such effects, however, are beyond the domain of applicability of the present framework. 

For applications of this framework to inflation, the impact on the curvature fluctuations as well as backreaction of curvature fluctuations on particle production is needed. The results from \cite{LopezNacir:2011kk,Green:2014xqa,Flauger:2016idt,Dias:2016slx} should provide useful guidance for such calculations. For reheating applications, the impact of different expansion histories on particle production and vice versa is needed. We hope that the technical results presented here and in \cite{Amin:2015ftc} will be useful as starting points for calculations related to observables.\footnote{In the present work we have focused on determining the total typical occupation number which is related to the sum of the eigenvalues of the transfer matrices. This is a ``basis independent" quantity, and was chosen for reasons of calculational convenience as well as usefulness in potential applications (for example, reheating studies on sub-horizon scales). In some applications, especially those related to superhorizon scales, such an occupation number need not have a useful physical meaning. Moreover, even on subhorizon scales, the quantities of interest may correspond to the most probable branching ratio $(n_a/n_b)_{\rm typ}$ of the individual occupation numbers $n_a$ for different fields or to other combinations of terms in the transfer matrices. Such quantities can in principle be evaluated, using this present work as a guide. We leave these calculations for future work.}

Finally, we note that while we were motivated by the desire to tackle particle production in the early universe, the new results presented here can be re-used  for the problem of current conduction in disordered wires (where the original motivation for the statistical framework came from). More generally, within the limits of the assumptions stated, our framework and results might be useful for analyzing systems described by a set of $\Nf$ coupled oscillators with $\Ns\gg 1$ localized and stochastic variations in their frequencies and couplings.

\section*{Acknowledgements}
We would like to thank  Anthony Chan, Matthew Foster, Daniel Baumann, Daniel Green, Mafalda Dias and Jonathan Frazer for helpful conversations. M.A.A. and M.A.G.G. acknowledge support from the US Department of Energy Grant DE-SC0018216. This work was completed at the Aspen Center for Physics, which is supported by National Science Foundation grant PHY-1066293.
\appendix

\section{Appendix}

\subsection{The Fokker-Planck Equation}\label{ap:FPd}

In this section we present a derivation of the Fokker-Planck equation (\ref{eq:FPlambda}), where the coefficient correlators are evaluated as functions of the matrices $\R(j)$ and $\R_{j+1}$. For this purpose we rewrite the Smoluchowski equation in the form
\beq
P_{\tau+\delta\tau}(\M)=\int d\mu(\M_1)\,d\mu(\M_2)\,P_{\tau}(\M_1)P_{\delta\tau}(\M_2)\,\delta(\M-\M_1\M_2)\,,
\eeq
or, equivalently, in terms of the $2\Nf^2+\Nf$ parameters $\blambda$ that characterize each transfer matrix,
\beq
P_{\tau+\delta\tau}(\blambda)=\int d\blambda_1\,d\blambda_2\,P_{\tau}(\blambda_1)P_{\delta\tau}(\blambda_2)\,\delta(\blambda-\tilde{\blambda}(\blambda_1,\blambda_2))\,,
\eeq
where the measure $d\blambda$ is suitably defined, and where $\tilde{\blambda}$ denotes the functional dependence of the parameters of a product of transfer matrix, i.e.~$\M(\tilde{\blambda})=\M_1(\blambda_1)\M_2(\blambda_2)$. Upon Fourier-transforming the previous equation, and introducing the change of variables $\tilde{\blambda}=\blambda_1+\delta \blambda $, we can write
\begin{align}
\tilde{P}_{\tau+\delta\tau}(\bomega) &\equiv \int d\blambda\,d\blambda_1\,d\blambda_2\,P_{\tau}(\blambda_1)P_{\delta\tau}(\blambda_2)\,\delta(\blambda-\tilde{\blambda}(\blambda_1,\blambda_2))\,e^{i\bomega\cdot\blambda}\\ \notag
& = \int d\blambda_1\,d\blambda_2\,P_{\tau}(\blambda_1)P_{\delta\tau}(\blambda_2)\,e^{i\bomega\cdot\tilde{\blambda}(\blambda_1,\blambda_2)}\\ \notag
& = \int d\blambda_1\,d\blambda_2\,P_{\tau}(\blambda_1)P_{\delta\tau}(\blambda_2)\,e^{i\bomega\cdot\blambda_1}e^{i\bomega\cdot\delta\blambda(\blambda_1,\blambda_2)}\\
& = \int d\blambda_1\,d\blambda_2\,P_{\tau}(\blambda_1)P_{\delta\tau}(\blambda_2)\,e^{i\bomega\cdot\blambda_1}\sum_{n=0}^{\infty}\frac{(i\bomega\cdot\delta\blambda)^n}{n!}\,.
\end{align}
Using the inverse Fourier transform,
\begin{align} \notag
P_{\tau+\delta\tau}(\blambda) &= \sum_{n=0}^{\infty} \frac{1}{n!} \int d\bomega \,d\blambda_1\,d\blambda_2\,P_{\tau}(\blambda_1)P_{\delta\tau}(\blambda_2)\,\frac{e^{i\bomega\cdot(\blambda_1-\blambda)}}{(2\pi)^{2\Nf^2+\Nf}}(i\bomega\cdot\delta\blambda)^n\\ \notag
&= \sum_{n=0}^{\infty} \frac{(-1)^n}{n!} \int d\blambda_1\,d\blambda_2\,P_{\tau}(\blambda_1)P_{\delta\tau}(\blambda_2)\,\left(\frac{\partial}{\partial\blambda}\right)^n\delta(\blambda- \blambda_1)\cdot\left(\delta \blambda \right)^n\\ \notag
&= \sum_{n=0}^{\infty} \frac{(-1)^n}{n!} \left(\frac{\partial}{\partial\blambda}\right)^n\cdot \int d\blambda_1\,d\blambda_2\,P_{\tau}(\blambda_1)P_{\delta\tau}(\blambda_2)\,\left( \delta \blambda \right)^n \delta(\blambda- \blambda_1)\\
&= \sum_{n=0}^{\infty} \frac{(-1)^n}{n!} \left(\frac{\partial}{\partial\blambda}\right)^n\cdot \Big[ \langle \left( \delta \blambda \right)^n \rangle_{-}\, P_{\tau}(\blambda) \Big]\,,
\end{align}
where
\beq
\langle \left( \delta \blambda \right)^n \rangle_{-} \equiv \lim_{\blambda_1\rightarrow \blambda}\int  d\blambda_2\,P_{\delta\tau}(\blambda_2)\,\left[ \delta \blambda(\blambda_1,\blambda_2) \right]^n\,.
\eeq
Upon division by $\delta\tau$, and integration with respect to the set of parameters that characterize $\v$ (which form a compact manifold, c.f.~(\ref{eq:Mparg})), the resulting equation corresponds to (\ref{eq:FPlambda}), where abusing notation we have also denoted by $P$ the probability density marginalized with respect to $\v$.

\subsection{Coefficients of the $\Nf=2$ FP Equation}\label{ap:nf2}

The list of increments $\delta\lambda_a$, needed to calculate the coefficients appearing in the $\Nf=2$ FP equation is provided in this appendix. Using eqns.~(\ref{eq:df1})-(\ref{eq:df4})), we have
\begin{subequations}  \label{1st}
\begin{align}
\delta{f}_\varrho^{(1)} & \, = \g^{(1)}_{\varrho,\varrho} ,\quad  \varrho \in \{1,2\}, \\ \displaybreak[0]
\delta{\theta}^{(1)}  & \, =\frac{2}{\Delta{f}} \Re{\left(\g^{(1)}_{2,1} e^{i \psi} \right)},  \\ \displaybreak[0]
\delta{\psi}^{(1)}  & \,  = \frac{1}{2} \Im\left(\frac{\tilde{\g}^{(1)}_{2,2}}{\tilde{f}_2} - \frac{{{\tilde{\g}}}^{(1)}_{1,1}}{\tilde{f}_1}\right)-\delta{\varphi}^{(1)} \cos{\theta}, \\
\delta{\phi}^{(1)}  & \, = -\frac{1}{2} \Im\left(\frac{{\tilde{\g}}^{(1)}_{2,2}}
{\tilde{f}_2} + \frac{{{\tilde{\g}}}^{(1)}_{1,1}}{\tilde{f}_1}\right), \\
\delta{\varphi}^{(1)} & \,  = \frac{2}{\Delta{f}}\Im{\left(\g^{(1)}_{2,1}e^{i \psi}\right)} \csc{\theta}\,,
\end{align}
\end{subequations}
where $\Delta f=f_1-f_2$. At next to leading order, we have
\begin{subequations}                    \label{2nd}
\begin{align}
\delta{f}_\varrho^{(2)} &\,=\g^{(2)}_{\varrho,\varrho} -(-1)^{\varrho} \frac{|\g^{(1)}_{2,1}|^2} {\Delta{f}}  ,\quad\varrho \in \{1,2\},  \\
\delta{\theta}^{(2)} &\, = \frac{2}{\Delta{f}} \Re{\left( \g^{(2)}_{2,1} e^{i \psi} \right)} +  \frac{1}{\Delta{f}}\left( \delta{f}_2^{(1)}-\delta{f}_1^{(1)} \right)\delta{\theta}^{(1)}+\frac{1}{4}\sin{2\theta}\left( \delta{\varphi}^{(1)} \right)^{2}\,,\\
\delta{\psi}^{(2)} &\, =  \frac{1}{2} \Im\left( \frac{\tilde{\g}^{(2)}_{2,2}}{\tilde{f}_2} - \frac{\tilde{{\g}}^{(2)}_{1,1}}{\tilde{f}_1} \right) -  \frac{1}{2} \left( \frac{f_1}{\tilde{f}_1^2}\delta{f}_1^{(1)} -\frac{f_2}{\tilde{f}_2^2} \delta{f}_2^{(1)} \right)\delta{\phi}^{(1)}\\
&\quad-\frac{1}{2} \left( \frac{f_1}{\tilde{f}_1^2}\delta{f}_1^{(1)} +\frac{f_2}{\tilde{f}_2^2} \delta{f}_2^{(1)} \right)\left(\hvp^{(1)}\cos{\theta}+ \delta\psi^{(1)}\right)\\
&\quad -\frac{1}{2}\left( \frac{\tilde{f}_1}{\tilde{f}_2} + \frac{\tilde{f}_2}{\tilde{f}_1} \right) \left(\frac{1}{4}\left[ \left(\hvp^{(1)}\right)^2 \sin^2{\theta} -\left( \hth^{(1)} \right)^2 \right]\sin{2\psi} +\frac{1}{2} \hvp^{(1)} \hth^{(1)} \sin{\theta} \cos{2\psi}\right)\\
&\qquad+\frac{1}{2}\sin{\theta}\delta{\theta}^{(1)} \delta{\varphi}^{(1)}-\cos{\theta}\delta{\varphi}^{(2)}, \\
\delta{\phi}^{(2)}   &\,  = -\frac{1}{2} \Im\left( \frac{\tilde{\g}^{(2)}_{2,2}}{\tilde{f}_2} + \frac{\tilde{{\g}}^{(2)}_{1,1}}{\tilde{f}_1} \right)-\frac{1}{2} \left( \frac{f_1}{\tilde{f}_1^2}\delta{f}_1^{(1)} +\frac{f_2}{\tilde{f}_2^2} \delta{f}_2^{(1)} \right) \hph^{(1)}\\
&\quad-\frac{1}{2} \left( \frac{f_1}{\tilde{f}_1^2}\delta{f}_1^{(1)} -\frac{f_2}{\tilde{f}_2^2} \delta{f}_2^{(1)} \right) \left(\hvp^{(1)}\cos{\theta}+ \delta\psi^{(1)}\right)\\
 &\quad+\frac{1}{2}\left( \frac{\tilde{f}_1}{\tilde{f}_2} - \frac{\tilde{f}_2}{\tilde{f}_1} \right) \left(\frac{1}{4}\left[ \left(\hvp^{(1)}\right)^2 \sin^2{\theta} -\left( \hth^{(1)} \right)^2 \right]\sin{2\psi} +\frac{1}{2} \hvp^{(1)} \hth^{(1)} \sin{\theta} \cos{2\psi}\right), \\
\delta{\varphi}^{(2)} &\,= \frac{2}{\Delta{f}} \Im{\left(\g^{(2)}_{2,1} e^{i \psi}\right)}
\csc{\theta}+ \frac{1}{\Delta{f}}\left( \delta{f}_2^{(1)}-\delta{f}_1^{(1)} \right) \delta{\varphi}^{(1)} -  \cot{\theta}\hth^{(1)}\hvp^{(1)},
\end{align}
\end{subequations}

The correlators $\langle \delta\lambda_a\rangle$ and $\langle \delta\lambda_a\delta\lambda_b\rangle$ that appear in the FP equation are listed next. The dependence of coefficients on the polar angle $\theta$ and the scatterings strengths can (almost) be encoded via the $\gamma,l$-functions defined below,
\begin{subequations} 
\begin{align} \label{eq:1stgamma}
\gamma_1(\theta) & \, =  2 \left[ 
  \sigma_1^2 \cos^4\left( \frac{\theta}{2} \right)
+ \sigma_2^2 \sin^4\left( \frac{\theta}{2} \right)
+ 4 \sigma_\perp^2 \sin^2\left( \frac{\theta}{2} \right) \cos^2\left( \frac{\theta}{2} \right)
\right], \\
\gamma_2(\theta) & \, =  2 \left[ 
  \sigma_1^2 \sin^4\left( \frac{\theta}{2} \right)
+ \sigma_2^2 \cos^4\left( \frac{\theta}{2} \right)
+ 4 \sigma_\perp^2 \sin^2\left( \frac{\theta}{2} \right) \cos^2\left( \frac{\theta}{2} \right)
\right], \\
\gamma_3(\theta) & \, = \frac{1}{4} \sin^2{\theta} \left( \sigma_1^2 + \sigma_2^2- 4\sigma_\perp^2 \right), \\
\gamma_4(\theta)&\,  = \frac{\sin{\theta}}{2} \left[ 
 \sigma_1^2 \cos^2 \left(\frac{\theta }{2}\right) 
-\sigma_2^2 \sin ^2\left(\frac{\theta }{2}\right)
-2 \sigma_\perp^2 \cos{\theta} \right], \\
\gamma_5(\theta)&\,  = \frac{\sin{\theta}}{2} \left[ 
 \sigma_1^2 \sin^2 \left(\frac{\theta }{2}\right) 
-\sigma_2^2 \cos^2 \left(\frac{\theta }{2}\right)
+ 2 \sigma_\perp^2 \cos{\theta} \right], \\
\gamma_6(\theta) &\,  = \frac{1}{2} \left[ \left( \sigma_1^2 + \sigma_2^2 \right) \sin^2{\theta}  + 4 \cos^2{\theta} \sigma_\perp^2  \right].
\end{align}
\end{subequations}
and
\begin{subequations}
\begin{align}
l_1(\theta) & \, =  2 \left[ \cos^2 \left(\frac{\theta}{2}\right) \sigma_1^2 + \sin^2 \left(\frac{\theta }{2}\right) \sigma_2^2 + \sigma_\perp^2 \right], \\
l_2(\theta) & \, =  2 \left[ \sin^2 \left(\frac{\theta}{2}\right) \sigma_1^2 + \cos^2 \left(\frac{\theta }{2}\right) \sigma_2^2 + \sigma_\perp^2 \right], \\
l_3(\theta) & \, =  2 \left[ \cos^4 \left(\frac{\theta}{2}\right) \sigma_1^2 + \sin^4 \left(\frac{\theta }{2}\right) \sigma_2^2 + \sin^2{\theta} \, \sigma_\perp^2 \right], \\ \label{eq:lastl}
l_4(\theta) & \, =  2 \left[ \sin^4 \left(\frac{\theta}{2}\right) \sigma_1^2 + \cos^4 \left(\frac{\theta }{2}\right) \sigma_2^2 + \sin^2{\theta} \, \sigma_\perp^2 \right], 
\end{align}
\end{subequations}
The correlations $\langle\delta\lambda_a\delta\lambda_b\rangle$ are given by 
\begin{subequations}  \label{1st-col-1}
\begin{align} 
\langle 
\delta f_1^{(1)} \delta f_1^{(1)} \rangle
= &\, \tilde{f}_1^2 \, \gamma_1(\theta), \\
\langle \delta f_1^{(1)} \delta f_2^{(1)} \rangle 
= &\, 2 \tilde{f}_1 \tilde{f}_2 \cos(2 \psi) \,\gamma_3(\theta), \\
\langle\delta f_1^{(1)} \delta \theta^{(1)} \rangle = &\, - \frac{2 \tilde{f}_1}{\Delta{f}} \left[\tilde{f}_1 + \tilde{f}_2 \cos(2 \psi)   \right] \gamma_4(\theta), \\
\langle \delta f_1^{(1)} \delta \psi^{(1)}  \rangle = &\, - \tilde{f}_{1}  \sin{(2\psi)} \left( \frac{f_{2}}{\tilde{f}_{2}} \gamma_3(\theta) -
2 \frac{\tilde{f}_2}{\Delta{f}}  \gamma_4(\theta) \cot{\theta} \right), \\
\langle \delta f_1^{(1)} \delta \varphi^{(1)}  \rangle = &\, - 2 \frac{\tilde{f}_1 \tilde{f}_2}{\Delta{f}} \sin(2 \psi) \gamma_4(\theta) \csc{\theta},\\
\langle\delta f_1^{(1)} \delta \phi^{(1)} \rangle = &\, \tilde{f}_1 \frac{f_2}{\tilde{f}_2} \sin(2\psi) \gamma_3(\theta), \\
\langle \delta f_2^{(1)}\delta f_2^{(1)}  \rangle = &\, \tilde{f}_2^2 \, \gamma_2(\theta), \\
\langle \delta f_2^{(1)} \delta \theta^{(1)}  \rangle = &\, - \frac{2 \tilde{f}_2}{\Delta{f}} \left[\tilde{f}_2 + \tilde{f}_1 \cos(2 \psi)   \right] \gamma_5(\theta), \\
\langle \delta f_2^{(1)} \delta \psi^{(1)}  \rangle = &\, - \tilde{f}_{2}  \sin{(2\psi)} \left( \frac{f_{1}}{\tilde{f}_{1}} \gamma_3(\theta) -
2 \frac{\tilde{f}_1}{\Delta{f}}  \gamma_5(\theta) \cot{\theta} \right), \\
\langle \delta f_2^{(1)} \delta \varphi^{(1)} \rangle = &\,- 2 \frac{\tilde{f}_1 \tilde{f}_2}{\Delta{f}} \sin(2 \psi) \gamma_5(\theta) \csc{\theta}, \\
\langle\delta f_2^{(1)} \delta \phi^{(1)}\rangle = &\, -\tilde{f}_2 \frac{f_1}{\tilde{f}_1} \sin(2\psi) \gamma_3(\theta),\\
\langle\delta \theta^{(1)} \delta \theta^{(1)}\rangle = &\, 2 \sigma_\perp^2 + \frac{1}{\Delta{f}^2} \left( \tilde{f}_1^2 + \tilde{f}_2^2 + 2\tilde{f}_1 \tilde{f}_2 \cos(2 \psi)  \right) \gamma_6(\theta), \\
\langle\delta \theta^{(1)} \delta \psi^{(1)}\rangle  = &\, \frac{1}{\Delta{f}}\left[ \frac{f_1 \tilde{f}_2}{\tilde{f}_1} \gamma_4(\theta)  +\frac{\tilde{f}_1 f_2}{\tilde{f}_2} \gamma_5(\theta)  \right] \sin(2 \psi) -\cos{\theta} \, \langle\delta \theta^{(1)} \delta \varphi^{(1)}\rangle, \\
\langle\delta \theta^{(1)} \delta \varphi^{(1)}\rangle = &\frac{2 \tilde{f}_1 \tilde{f}_2 \sin(2\psi) }{\Delta{f}^2} \, \gamma_6(\theta) \csc{\theta}, \\ 
\langle\delta \theta^{(1)} \delta \phi^{(1)}\rangle=&\, \frac{1}{4\Delta f}\left[
\left(f_1 \frac{\tilde{f}_2}{\tilde{f}_1}-f_2 \frac{\tilde{f}_1}{\tilde{f}_2}\right)(\sigma_1^2-\sigma_2^2)\sin\theta+4\left(f_1 \frac{\tilde{f}_2}{\tilde{f}_1}+f_2 \frac{\tilde{f}_1}{\tilde{f}_2}\right)\cot\theta \gamma_3(\theta)\right]\sin (2\psi)\,\\ \notag
\langle\delta \psi^{(1)} \delta \psi^{(1)}\rangle  = &\, \frac{1}{4} \left[ \left(\frac{f_1}{\tilde{f}_1}\right)^2 \gamma_1(\theta) + \left(\frac{f_2}{\tilde{f}_2}\right)^2 \gamma_2(\theta) \right] -\frac{f_1 f_2}{\tilde{f}_1 \tilde{f}_2} \gamma_3(\theta) \cos(2 \psi) + \cos^2{\theta} \left( \sigma_1^2 + \sigma_2^2 \right)\\
&\,+ 2\sin^2{\theta}\sigma_\perp^2 -\cos{\theta} \left( 2 \, \langle\delta \psi^{(1)} \delta \varphi^{(1)}\rangle+ \cos{\theta}\,\langle\delta \varphi^{(1)} \delta \varphi^{(1)}\rangle  \right)\,, \\ \notag
\langle\delta \psi^{(1)} \delta \varphi^{(1)}\rangle 
= &\, -\frac{1}{\Delta{f}}\left( \frac{f_1 \tilde{f}_2}{\tilde{f}_1} \gamma_4(\theta) + \frac{\tilde{f}_1 f_2}{\tilde{f}_2} \gamma_5(\theta) \right) \csc{\theta} \, \cos(2\psi) + \frac{f_1+f_2}{4 \Delta{f}}\left(\sigma_1^2 - \sigma_2^2\right) \\ 
&\,+ \left[\frac{5}{4}\left(\sigma_1^2 + \sigma_2^2\right) -3 \sigma_\perp^2\right] \cos{\theta}  -\cos{\theta} \, \langle\delta \varphi^{(1)} \delta \varphi^{(1)}\rangle, \\ \notag
\langle\delta \psi^{(1)} \delta \phi^{(1)}\rangle = &\, \frac{1}{2}\left(\frac{f^2_1}{{\tilde f}^2_1}-\frac{f^2_2}{{\tilde f}^2_2}\right)\left[\frac{1}{2}\left(\sigma_1^2+\sigma_2^2\right)-\gamma_3(\theta)\right]\\
&\,+\left[1+\frac{1}{4}\left(\frac{f^2_1}{{\tilde f}^2_1}+\frac{f^2_2}{{\tilde f}^2_2}\right)\right](\sigma_1^2-\sigma_2^2)\cos\theta-\langle\delta\varphi^{(1)}\delta\phi^{(1)} \rangle\cos\theta,\\
\langle\delta \varphi^{(1)} \delta \varphi^{(1)}\rangle = &\, \left( \sigma_1^2 + \sigma_2^2 \right) + 2 \cot^2{\theta} \,\sigma_\perp^2 + \frac{1}{\Delta{f}^2} \left[ \tilde{f}_1^2 + \tilde{f}_2^2 - 2 \tilde{f}_1 \tilde{f}_2 \cos(2 \psi) \right] \gamma_{6} (\theta) \csc^2\theta, \\ \notag
\langle\delta \varphi^{(1)} \delta \phi^{(1)}\rangle 
= &\, \frac{1}{\Delta f}\left[f_1+f_2-\left({\tilde f}_1 \frac{f_2}{{\tilde f}_2}+{\tilde f}_2 \frac{f_1}{{\tilde f}_1}\right)\cos(2\psi)\right]\cot\theta{\,\rm csc\,}\theta \gamma_3(\theta) \\
&\, +\frac{1}{4\Delta f}(\sigma_1^2-\sigma_2^2)\left[5(f_1-f_2)+\left({\tilde f}_1 \frac{f_2}{{\tilde f}_2}-{\tilde f}_2 \frac{f_1}{{\tilde f}_1}\right)\cos(2\psi)\right]\,,\\ 
\langle\delta \phi^{(1)} \delta \phi^{(1)}\rangle = &\, \left( \sigma_1^2 + \sigma_2^2 \right)\left(1+\frac{1}{4}\left(\frac{f^2_1}{{\tilde f}^2_1}+\frac{f^2_2}{{\tilde f}^2_2}\right)\right)
+\left( \sigma_1^2- \sigma_2^2 \right)\frac{1}{4}\left(\frac{f^2_1}{{\tilde f}^2_1}-\frac{f^2_2}{{\tilde f}^2_2}\right)\cos \theta\\
&\,-\frac{1}{2}\left[\frac{f^2_1}{{\tilde f}^2_1}+\frac{f^2_2}{{\tilde f}^2_2}-2\frac{f_1}{{\tilde f}_1}\frac{f_2}{{\tilde f}_2}\cos(2\psi)\right]\gamma_3(\theta),
\end{align}
\end{subequations}
while the expectation values $\langle\delta\lambda_a\rangle$ correspond to
\begin{subequations}  \label{2nd-col-1}
\begin{align}
\langle\delta{f}_{1}^{(2)}\rangle &\, = \frac{1}{\Delta{f}} \left[ \tilde{f}_1^2 l_1(\theta) -(f_1 f_2 -1) l_3(\theta)  \right], \\
\langle\delta{f}_{2}^{(2)}\rangle &\, = -\frac{1}{\Delta{f}} \left[ \tilde{f}_2^2 l_2(\theta) -(f_1 f_2 -1) l_4(\theta)  \right], \\ \notag
\langle\delta{\theta}^{(2)}\rangle &\, = -\frac{(f_1+f_2)}{\Delta{f}} \left(\sigma _1^2-\sigma _2^2\right)  \sin{\theta}  -\frac{1}{2} \sin (2 \theta ) \left(\sigma _1^2+\sigma _2^2 -3\sigma_\perp^2 \right)\\
&\quad + \frac{1}{\Delta f} \left(\langle \delta f_2^{(1)}\delta\theta^{(1)}\rangle-\langle \delta f_1^{(1)}\delta\theta^{(1)}\rangle\right) + \frac{1}{4} \sin{(2 \theta)} \langle\delta\varphi^{(1)}\delta\varphi^{(1)}\rangle, \\ \notag
\langle\delta{\psi}^{(2)}\rangle &\, = \frac{1}{8} \left(\frac{\tilde{f}_1}{\tilde{f}_2} + \frac{\tilde{f}_2}{\tilde{f}_1}  \right)
\left( \sigma_1^2 + \sigma_2^2- 2 \sigma_\perp^2 \right)\sin^2{\theta} \sin{(2\psi)}-   \frac{1}{2} \left( \frac{f_1}{\tilde{f}_1^2}\langle\delta{f}_1^{(1)}\delta{\phi}^{(1)}\rangle - \frac{f_2}{\tilde{f}_2^2} \langle\delta{f}_2^{(1)}\delta{\phi}^{(1)}\rangle \right)
\\ \notag
&\quad- \frac{1}{2} \left( \frac{f_1}{\tilde{f}_1^2}\langle\delta{f}_1^{(1)}\delta\psi^{(1)}\rangle+ \frac{f_2}{\tilde{f}_2^2} \langle\delta{f}_2^{(1)}\delta\psi^{(1)}\rangle \right)-\frac{1}{2} \left( \frac{f_1}{\tilde{f}_1^2}\langle\delta{f}_1^{(1)}\delta\varphi^{(1)}\rangle+ \frac{f_2}{\tilde{f}_2^2} \langle\delta{f}_2^{(1)}\delta\varphi^{(1)}\rangle \right)\cos{\theta}\\ \notag
&\quad -\frac{1}{2}\left(\frac{\tilde{f}_1}{\tilde{f}_2}-\frac{\tilde{f}_2}{\tilde{f}_1}\right) \left[\frac{1}{4}\left( \langle\hvp^{(1)}\hvp^{(1)}\rangle \sin^2{\theta} -\langle \hth^{(1)}\hth^{(1)}\rangle \right)\sin{2\psi} +\frac{1}{2} 
\langle\hvp^{(1)} \hth^{(1)}\rangle \sin{\theta} \cos{2\psi}\right] \\
&\quad+\frac{1}{2}\sin{\theta} \, \langle\delta{\theta}^{(1)}\delta{\varphi}^{(1)}\rangle-\cos{\theta}\langle\delta{\varphi}^{(2)}\rangle, \\
\langle\delta{\varphi}^{(2)}\rangle &\, =  \frac{1}{\Delta f}  \left(\langle\delta f_2^{(1)}\delta{\varphi}^{(1)}\rangle-\langle\delta f_1^{(1)}\delta{\varphi}^{(1)}\rangle\right)  -  \cot{\theta} \, \langle\hth^{(1)}\hvp^{(1)}\rangle\,, \\
\langle\delta{\phi}^{(2)}\rangle &\, =  0\,.\end{align}
\end{subequations}

\subsection{Numerical Method}\label{ap:nm}

\begin{figure}[t!]
\centering
    \scalebox{0.70}{\includegraphics{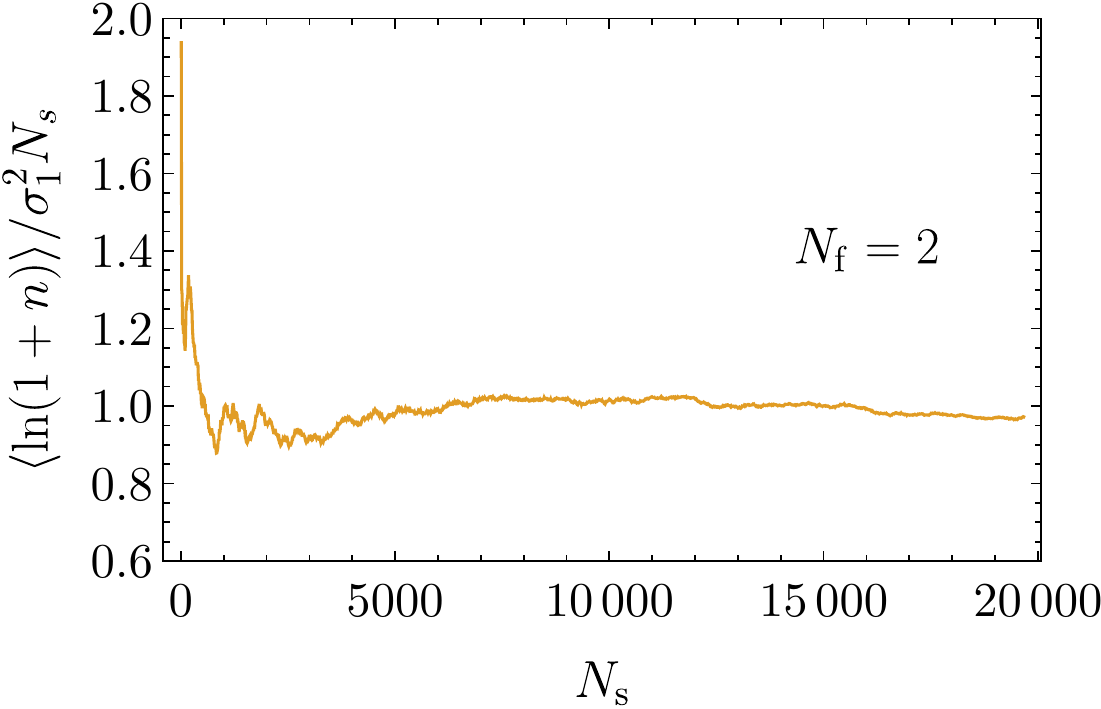}}
    \hfill
    \scalebox{0.70}{\includegraphics{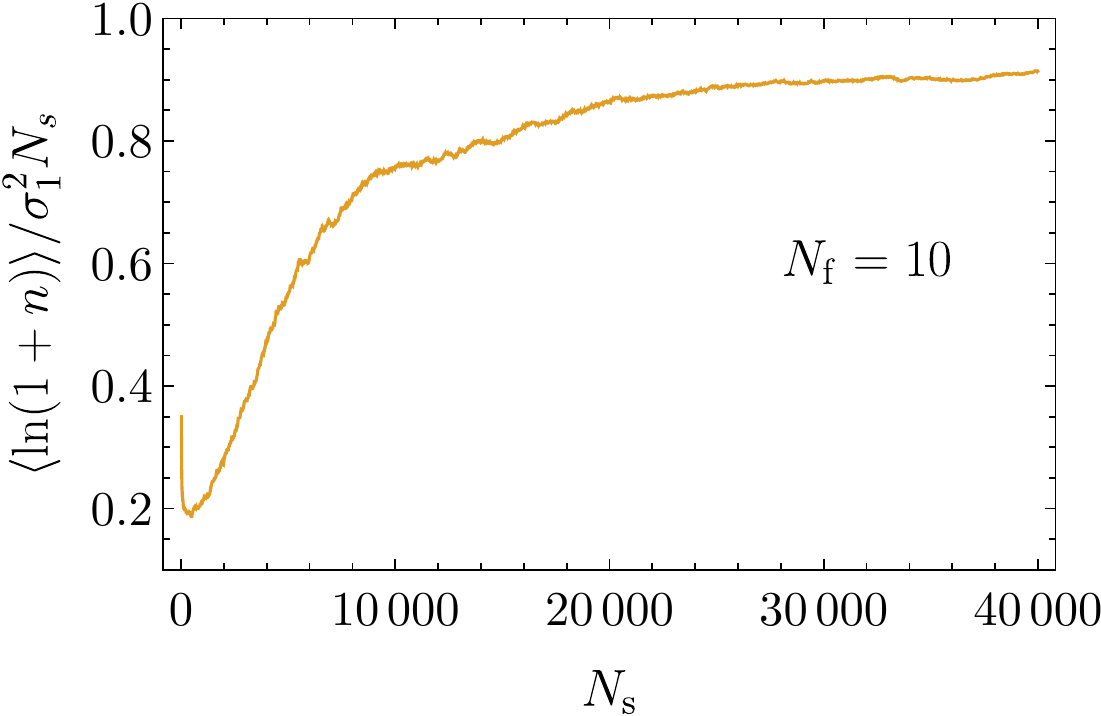}}
    \caption{The numerically calculated particle production rate as a function of the number of scatterings, for a particular realization of the scattering locations and strengths. For both panels $(\sigma_2/\sigma_1)^2=10^{-3} $ and $(\sigma_{\perp}/\sigma_1)^2=10^{-2}$. Left: the production rate for two interacting fields. Right: the production rate with ten fields.}
    \label{fig:N102P}
\end{figure}

The transfer matrix approach described in section~\ref{sec:TMA} can be translated in a straightforward way to a simple numerical method to compute the occupation number after a certain number of scatterings; the occupation numbers $n_a(j)$ are obtained by keeping track of the iterated multiplication of matrices and evaluating its trace at each step. However, as it is shown in figure~\ref{fig:N102P}, in order to reach the stationary late-time limit for the particle production rate, for which our analytical results are valid, an enormous amount of calculations need to be considered as $\Nf$ increases. Since such a computation using the naive approach would almost inevitably lead to numerical overflow, a better algorithm is needed.

Recall that the solution for the Fourier mode of the field $\chi^a$ after $j$ scattering events is given by (\ref{equ:chi}), where the amplitudes of the exponential terms for all fields can be arranged in the coefficient vector
\beq
\boldsymbol{\Psi}(j) \equiv \begin{pmatrix} \vec{\beta}_{j} \\ \vec{\alpha}_{j} \end{pmatrix}\,.
\eeq
Note first that
\beq
\ln\left(|\boldsymbol{\Psi}(j)|^2\right) = \ln\left(|\vec{\beta}_{j}|^2 + |\vec{\alpha}_{j}|^2\right) = \ln\left(2n(j)+\Nf \right)  \xrightarrow[]{ \Ns\gg 1 } \ln\left(2n(j) \right)\,,
\eeq
and $|\boldsymbol{\Psi}(0)|^2=\Nf$. Therefore, if we split the chain of $\Ns$ scatterings into $m$ blocks containing each $N_m$ interactions, $\Ns=m\times N_m$, the typical particle number rate in the $\Ns\gg 1$ limit can be approximated as
\begin{align} \notag
\Ns^{-1}\langle \ln(1+n)\rangle &\simeq \Ns^{-1}\langle \ln( |\boldsymbol{\Psi}(\Ns) |^2)\rangle\\ \notag
& = \Nf\Ns^{-1}\left\langle \ln\left(\prod_{j=1}^{m}\frac{|\boldsymbol{\Psi}(jN_m) |^2}{|\boldsymbol{\Psi}((j-1)N_m) |^2}\right)\right\rangle\\ 
& = \Nf\Ns^{-1}\left\langle \sum_{j=1}^{m} \ln\left(\frac{|\boldsymbol{\Psi}(jN_m) |^2}{|\boldsymbol{\Psi}((j-1)N_m) |^2}\right)\right\rangle\,,
\end{align}
i.e.~we compartmentalize the matrix multiplication; the argument of the logarithm is identified with the amount that the arbitrary unit vector is scaled during each block. At the start of each $j^{\rm th}$ block we  re-normalize the outcome of the previous block, $|\boldsymbol{\Psi}((j-1)N_m) |^2=1$, and apply the transfer matrix $N_m$ times to obtain $|\boldsymbol{\Psi}(jN_m) |^2$; the procedure is then repeated for the $(j+1)^{\rm th}$ block and the results are added. Upon finding the particle production rate for a particular realization of the scattering locations and strengths, the procedure is repeated for several realizations, and the expectation value returns the desired  mean particle production rate.

\subsection{Haar average of $\Fo_{\Omega}$}\label{ap:Fav}

In the notation of section~\ref{subsec:Ncorr}, the Haar measure of the group $SU(N)$ can be explicitly written as~\cite{TilmaSudarshanSUN}
\beq
d\mu\big(SU(N)\big) = K_{SU(N)}\,d\alpha_{N^2-1}\cdots d\alpha_1\,,
\eeq
where the kernel of the measure is given by
\beq
K_{ SU(N)} = \prod_{N\geq m\geq 2}\left(\prod_{2\leq k\leq m}{\rm Ker}(k,j(m))\right)\,,
\eeq
with
\beq
{\rm Ker}(k,j(m)) = \begin{cases}
\sin(\alpha_{2+j(m)})\,, & k=2\\
\cos(\alpha_{2(k-1)+j(m)}/2)^{2k-3}\sin(\alpha_{2(k-1)+j(m)}/2)\,, & 2<k<m\\
\cos(\alpha_{2(m-1)+j(m)}/2)\sin(\alpha_{2(m-1)+j(m)}/2)^{2m-3}\,, & k=m
\end{cases}
\eeq
and
\beq
j(m) = \begin{cases}
0 \,, & m=N\\
\sum_{0\leq l\leq N-m-1} 2(m+l)\,,  & m\neq N\,.
\end{cases}
\eeq
For the subset of angles that parametrize $\Fo_{\Omega}$, one can immediately check that
\beq
K_{ SU(N)} \propto \sin(\alpha_2)\cos^3(\alpha_4/2)\sin(\alpha_4/2) \cdots\cos^{2N-5}(\alpha_{2(N-2)}/2)\sin (\alpha_{2(N-2)}/2)\,.
\eeq
Therefore, the expectation value $\langle\Fo_{\Omega}\rangle_{\rm Haar}$ can be calculated inductively using the defining expression (\ref{eq:Fodef}), as follows:
\begin{alignat*}{2}
&\alpha_2:\qquad &&\frac{\int [\cos^4(\alpha_2/2)+\sin^4(\alpha_2/2)]\sin(\alpha_2)\,d\alpha_2}{\int  \sin(\alpha_2)\,d\alpha_2}\;=\;\frac{2}{3}\,,\\
&\alpha_4:\qquad &&\frac{\int [\frac{2}{3}\cos^4(\alpha_4/2)+\sin^4(\alpha_4/2)]\cos^3(\alpha_4/2)\sin(\alpha_4/2)\,d\alpha_4}{\int  \cos^3(\alpha_4/2)\sin(\alpha_4/2)\,d\alpha_4}\;=\;\frac{1}{2}\,,\\
&\vdots\\
&\alpha_{2(m-1)}:\  &&\frac{\int [\frac{2}{m}\cos^4(\alpha_{2(m-1)}/2)+\sin^4(\alpha_{2(m-1)}/2)]\cos^{2m-3}(\alpha_{2(m-1)}/2)\sin(\alpha_{2(m-1)}/2)\,d\alpha_{2(m-1)}}{\int  \cos^3(\alpha_{2(m-1)}/2)\sin(\alpha_{2(m-1)}/2)\,d\alpha_{2(m-1)}}\\
& && \qquad \;=\;\frac{2/(m^2-1)}{1/(m-1)} \;=\; \frac{2}{m+1}\,,\\
&\vdots
\end{alignat*}
As for the last step we have $\alpha_{2(N-2)}$, we obtain $\langle\Fo_{\Omega}\rangle_{\rm Haar}=2/N$.

\addcontentsline{toc}{section}{References}
\bibliographystyle{utphys}
\bibliography{Refs}

\end{document}